%% file: massive_cfht.tex
\documentclass[fleqn,usenatbib]{mnras}

\usepackage[T1]{fontenc}
\usepackage{ae,aecompl}

\usepackage[utf8]{inputenc} % Let text accents work
\usepackage{graphicx}	% Including figure files
\usepackage{amsmath}	% Advanced maths commands
\usepackage{amssymb}	% Extra maths symbo
\usepackage{longtable}	% Allow for multi-page tables
\usepackage{catchfile}
\usepackage{caption}
\usepackage[dvipsnames]{xcolor}
\usepackage{tablefootnote}

\newcommand{\atlas}{ATLAS\textsuperscript{3D}\,}
\newcommand{\Rapp}{R_{e,\mathrm{app}}}

\newcommand{\magsec}{\ensuremath{\hbox{mag}\,\,\hbox{arcsec}^{-2}}}
\newcommand{\mbh}{$M_{\rm BH}$}

\title[MASSIVE CFHT Photometry]{The MASSIVE Survey. XVIII. Deep Wide-Field $K$-band Photometry and Local Scaling Relations for Massive Early-Type Galaxies}

\author[M. E. Quenneville et al.]{Matthew E.~Quenneville,$^{1,2}$\thanks{E-mail: mquenneville@berkeley.edu} 
John P.~Blakeslee,$^3$ 
Chung-Pei Ma,$^{1,2}$
Jenny E.~Greene,$^4$
\newauthor
Stephen D.~J.\ Gwyn,$^5$
Stephanie Ciccone,$^6$ and
Blanka Nyiri$^7$ \vspace{6pt}\\
$^{1}$Department of Physics, University of California, Berkeley, CA 94720, USA\\
$^{2}$Department of Astronomy, University of California, Berkeley, CA 94720, USA\\
$^{3}$NSF's NOIRLab, 950 N. Cherry Avenue, Tucson, AZ 85719, USA\\
$^{4}$Department of Astrophysical Sciences, Princeton University, Princeton, NJ 08544, USA\\
$^{5}$NRC Herzberg Astronomy and Astrophysics, 5071 West Saanich Road, Victoria, BC, V9E 2E7, Canada\\
$^{6}$Department of Physics, University of Guelph, Guelph, ON, N1G 2W1, Canada\\
$^{7}$Department of Physics and Astronomy, University of Waterloo, Waterloo, ON N2L 3G1, Canada
}

\date{Accepted XXX. Received YYY; in original form ZZZ}

\pubyear{2022}

\begin{document}
\label{firstpage}
\pagerange{\pageref{firstpage}--\pageref{lastpage}}
\maketitle

% Abstract of the paper
\begin{abstract}
We present wide-field, deep $K$-band photometry of 98 luminous early-type galaxies (ETGs) from the MASSIVE survey based on observations taken with the WIRCam instrument on the Canada-France-Hawaii Telescope. Using these images, we extract accurate total $K$-band luminosities ($L_K$) and half-light radii ($R_e$) for this sample of galaxies. We use these new values to explore the size-luminosity and Faber-Jackson relations for massive ETGs. Within this volume-limited sample, we find clear evidence for curvature in both relations, indicating that the most luminous galaxies tend to have larger sizes and smaller velocity dispersions than expected from a simple power-law fit to less luminous galaxies. Our measured relations are qualitatively consistent with the most massive elliptical galaxies forming largely through dissipationless mergers. When the sample is separated into fast and slow rotators, we find the slow rotators to exhibit similar changes in slope with increasing $L_K$, suggesting that low-mass and high-mass slow rotators have different formation histories. The curvatures in the $R_e-L_K$ and $\sigma-L_K$ relations cancel, leading to a relation between dynamical mass and luminosity that is well described by a single power-law: $R_e\sigma^2 \propto {L_K}^b$ with $b\approx 1.2$. This is consistent with the tilt of the fundamental plane observed in lower mass elliptical galaxies.
\end{abstract}

% Select between one and six entries from the list of approved keywords.
% Don't make up new ones.

%%%%%%%%%%%%%%%%%%%%%%%%%%%%%%%%%%%%%%%%%%%%%%%%%%

%%%%%%%%%%%%%%%%% BODY OF PAPER %%%%%%%%%%%%%%%%%%

\section{Introduction}

By studying the properties of nearby massive early-type galaxies (ETGs), we can learn
about their evolutionary histories. The growth history of these galaxies can leave
measurable impacts on their observed properties. Many of these properties are found to be
strongly correlated. For instance, the size-luminosity (SL) relation describes the correlation between a
galaxy's projected half-light radius ($R_e$) and its total luminosity ($L$), and the Faber-Jackson
(FJ) relation describes a correlation between a galaxy's velocity dispersion along the line-of-sight ($\sigma$) and total luminosity \citep[e.g.,][]{Faber1987,Robertson2006,Bernardietal2011,Cappellarietal2013a}. 
When all three of these parameters are considered simultaneously, elliptical galaxies are distributed on a single 3-dimensional power-law relation with relatively little scatter, known as the fundamental plane \citep[e.g.,][]{Dressleretal1987,DjorgovskiDavis1987}. 

Curvature has been observed in several scaling relations \citep[e.g.,][]{OegerleHoessel1991,Lauer2007a,vonderLindenetal2007,HydeBernardi2009,Samiretal2020} and may indicate changes in the relative influence of different formation mechanisms with galaxy luminosity. One example of this is the role of dissipation in galaxy mergers. Dissipation is thought to play a larger role in the merger history of lower mass, rotation dominated ellipticals and a smaller role for more massive, slow-rotating ellipticals~\citep[e.g.,][]{Benderetal1992,KormendyBender1996,Faberetal1997,Naabetal2006,Krajnovicetal2018}. In dissipation-less mergers, the pre-merger trajectories may leave lasting signatures within the resulting galaxy~\citep[e.g.,][]{NaabBurkert2003,BK2005,Jesseitetal2005, Jesseitetal2009}. These effects can, in turn, impact scaling relations for these galaxies. For example, \citet{Boylan-Kolchin2006} and \citet{Bernardietal2011} put forward arguments suggesting that pre-merger orbit trajectories may leave an imprint in the form of curvatures in the SL and FJ relations for massive ellipticals. 

There is also evidence that the outer envelopes of massive ellipticals grow along with the dark matter halos that they inhabit, such that at fixed stellar mass at radii ${\lesssim\,}10$ kpc, the outer envelope increases in mass along with the group/cluster mass \citep{Huangetal2018, Huangetal2020}. This effect is reported in weak lensing observations, which show that indeed the envelope mass measured between 50-100 kpc correlates with halo mass as tightly as standard richness measures \citep{Huangetal2022}. A similar effect is seen in the scaling of the total number of halo globular clusters, which tend to have a more extended distribution than the stellar light, with the mass of the dark matter halo \citep[e.g.,][]{Blakeslee1999,Alamoetal2013,Hudsonetal2014,Bastienetal2020}.
A change in slope of the FJ or SL relation may reflect growth of this outer envelope, which is likely driven by late-time accretion and does not impact the galaxy core.

In order to determine the slopes and intercepts of these scaling relations, accurate and precise measurements are needed of the velocity dispersions, half-light radii, and total luminosities. 
The MASSIVE survey is an ongoing effort to measure and characterize the properties of the most massive nearby ETGs~\citep{Maetal2014}. The full sample comprises 116 galaxies with absolute $K$-band magnitude $M_K <-25.3$ mag as measured in the 2MASS extended source catalog, corresponding to stellar
masses $M^*\gtrsim 10^{11.5} M_\odot$. The survey is volume limited within a distance of about 100~Mpc in the northern sky.
We have obtained wide-field spectroscopic data from the McDonald Mitchell IFS
as well as high spatial-resolution spectroscopic data from Gemini GMOS
to study the stellar kinematics of these galaxies. Spatially-resolved kinematic measurements are analyzed for a large sample of MASSIVE galaxies in \citet{Vealeetal2017a, Vealeetal2017b, Vealeetal2018, Eneetal2018, Eneetal2019} and \citet{Eneetal2020}. 

In this paper, we turn to the photometric properties of these galaxies and measure the half-light radius $R_e$  and total luminosity $L_K$ for 98 galaxies in the MASSIVE survey for which we have obtained deep, wide-field $K$-band imaging. $K$-band imaging already exists for the
MASSIVE galaxies from the 2MASS extended source catalog (XSC). However, total magnitudes from the XSC have been found to be systematically too faint. The half-light radii from the 2MASS XSC tend to
be systematically smaller than other estimates, perhaps due to the underestimate of the total luminosity. This may be due to a combination of extrapolation from insufficiently deep photometry \citep[e.g.,][]{Lauer2007a} and systematic issues in the 2MASS analysis pipeline~\citep{SchombertSmith2012}.  
These issues are mitigated in our imaging data, which reach 2.5 mag deeper than 2MASS.

Imaging in the $K$-band is particularly useful since it accurately traces the stellar populations within ETGs, minimizes extinction due to dust, and enables uniform calibration using 2MASS. In addition to the size-luminosity and FJ relations, accurate total luminosities and half-light radii are key to many other science goals of the MASSIVE survey. Half-light radii provide a natural scale for each galaxy and are used in many analyses, including studies of stellar dynamics \citep{Eneetal2019} and stellar population and initial mass function \citep{Guetal2022}. Total luminosities are needed to study the supermassive black hole (SMBH) and host galaxy \mbh-$L$ relation, one of the most commonly used local SMBH scaling relations \citep[e.g.,][]{McConnellMa2013, Kormendy2013}.

In Section~\ref{sec:reduction}, we describe the CFHT WIRCam observations, the data reduction process, and the method for determining total luminosities and half-light radii from the reduced images. In Section~\ref{sec:photometric_params}, we compare the resulting parameter values to those from
2MASS, and demonstrate the systematic bias in $K$-band magnitudes and $R_e$ from 2MASS. We discuss measurement errors and outline our fitting procedure in Section~\ref{sec:scalingrelations}, before analyzing the SL and FJ relations in Section~\ref{sec:SL} and Section~\ref{sec:FJ} respectively. Section~\ref{sec:virial} explores the relationship between a dynamical mass estimator and total lumionsity. 

\section{Observations and Reductions}
\label{sec:reduction}

\subsection{CFHT WIRCam Observations}

The selection of galaxies for the MASSIVE survey is described in detail by
\citet{Maetal2014} and is based on 2MASS photometry combined with distances estimated from
the 2MASS galaxy redshift survey \citep{Huchra2012}.  To obtain improved estimates of
the photometric and structural parameters for the MASSIVE sample, we targeted the galaxies
for deep wide-field near-infrared imaging with the Wide-field InfraRed Camera
\citep[WIRCam;][]{WIRCam2004} on the Canada-France-Hawaii Telescope (CFHT).  We chose to
use the $K$ band for these observations because it traces the old populations that
make up most of the stellar mass in these galaxies, and it minimizes dust extinction.
Priority was given to the subset of 72 MASSIVE galaxies with absolute $M_K <-25.5$\,mag,
as these were also the priorty targets for the integral field spectroscopy.

The observations presented here were conducted by CFHT staff in Queued Service Observation mode over a series of semesters from late 2014 to early 2017.  The focal plane of WIRCam contains four HAWAII-2RG detectors imaged at a pixel scale of 0\farcs307~pix$^{-1}$, so that each detectors covers $10\farcm4\times10\farcm4$. They are arranged in a square mosaic with $\sim\,$0\farcm6 gaps between the detectors. The full field of view thus spans approximately 21\farcm5 square.

We used the ``WIRCam Dithering Pattern 5'' (WDP5) sequence that successively places
the target on each of the four detectors of the mosaic, and then steps through the
sequence five times, ensuring small offsets of about 1\arcsec\ between subsequent
placements of the target on the same detector, for a total of 20 exposures. For the first
two semesters of the program, we executed this pattern twice in succession with individual
exposures of 20\,s, followed by a WDP3 sequence (stepping through all four chips 3 times)
with exposures of 10\,s to avoid saturation. This gave total on-target exposure times of
920\,s for very deep images.  With overheads, the executed time was 24 minutes per
target, not including slew.

However, by analyzing the images, we found that this amount of exposure was excessive for
our purpose.  Tests showed the results were not significantly affected when using half of
the exposure stack, as systematic effects in the sky estimation become dominant.  In
addition, we found that in good seeing conditions, the centers of some galaxies could
saturate even in 10\,s exposures (these were later reobserved).  We therefore adopted a
revised observing strategy that consisted of a single ``long'' WDP5 sequence with 20\,s
exposures followed by another ``short'' sequence with 3\,s exposures to ensure that none
of the galaxies saturated in the center.  The resulting total exposure time was thus
460\,s per galaxy.  Under typical conditions, this approach yielded a 3-$\sigma$ surface
brightness limit of $\mu_K\approx\,$23.0 AB\,\magsec, roughly 2.5~mag fainter than 2MASS.
Repeat observations showed that there were no systematic differences between the results
obtained with the original and revised observing strategies.

In all, we obtained high quality imaging for 98 MASSIVE survey galaxies, but for the luminous galaxy pair NGC~545 and NGC~547, our standard photometric analysis procedure did not yield reliable results because their isophotes are so strongly overlapping. These two galaxies are excluded below.

\subsection{Image Processing}

Standard detrending of the WIRCam exposures was performed by the IDL Interpretor of WIRCam
Images (`I`iwi) processing pipeline at
CFHT.\footnote{https://www.cfht.hawaii.edu/Instruments/Imaging/WIRCam/~\\IiwiVersion1Doc.html}
`I`iwi identifies and flags saturated pixels, corrects the pixel intensities for nonlinearity
effects, performs bias and dark current subtraction, divides by the normalized flat field, and then
masks the known bad pixels.  It also performs initial sky subtraction using offset fields; the sky
level estimation is improved during the stacking process.

The sets of detrended exposures for each galaxy observation were then processed with the WIRWolf image stacking pipeline \citep{Gwyn2014}. WIRWolf performs automatic photometric and astrometric calibration by matching the detected sources in each WIRCam exposure against the 2MASS catalog data.  After this initial iteration, it then matches the stellar magnitudes for each exposure to a master catalogue generated from the full set of images in the stack. This procedure results in an internal photometric accuracy typically better than 0.003~mag for each exposure relative to all the
others.  The calibrated images are resampled onto a common grid of pixel size 0\farcs3 and stacked using SWarp \citep{Terapix2002}.

After stacking, WIRWolf again compares the point source magnitudes with 2MASS to 
determine the final photometric calibration, which is then recorded in the header of the stacked image in terms of the natural AB system of WIRCam.  For the current analysis, we convert from AB to Vega magnitudes by subtracting 1.88~mag, and then transform to the 2MASS system using measured
relations,\footnote{https://www.cadc-ccda.hia-iha.nrc-cnrc.gc.ca/en/wirwolf/} which give a small
offset of 0.02~mag between the two systems at the $(H{-}K)_\mathrm{2MASS}\approx0.29\pm0.02$ color
of elliptical galaxies \citep[e.g.,][]{Carteretal2009}.  
The magnitudes are corrected for Galactic extinction using the values and extinction ratio given by \citet{Maetal2014}. Thus, the final $K$ magnitudes presented
here are on the standard Vega-based 2MASS system, corrected for extinction, and have a typical calibration accuracy of 0.03~mag.

In some of the images, a cross-shaped pattern is apparent, centered on a star that was used for
guiding the telescope during the observation. This pattern indicates a deficit of light within the
columns and rows containing the star. In these cases, we mask the affected rows and columns within
the image. In a few cases, the galaxy itself was used for guiding, resulting in a cross pattern on
the galaxy nucleus. These observations could not be used, and we reobserved the galaxies in
subsequent runs, explicitly instructing the software to avoiding guiding on sources within a
30\arcsec\ radius of the galaxy center.  In addition, some very bright stars can producing a
bleeding effect, giving an excess of light along the entire column containing the star. In these
cases, we also masked the affected column.

In addition to the above problems relating to stars in the images, vertical bands of variable bias levels appeared in the WIRCam data in 2016A, and CFHT implemented a correction by subtracting the median of each column. However, due to the extended nature of our targets, this did not work well for our data and resulted in significant deficits of light above and below the galaxy.  We corrected for this by aggressively masking the main galaxy and all other foreground sources above the local background, and then subtracting the median values from each column determined near the top and bottom of the image. This essentially removed the correction applied by CFHT.  We then requested the observatory to process all subsequent data without applying that correction.  The vertical bias bands tend to average out in the stacked image.

\section{Galaxy Parameter Measurements}
\label{sec:photometric_params}

\subsection{Curve of Growth Analysis}
\label{sec:archangel}

\begin{figure*}
	\includegraphics[width=\textwidth]{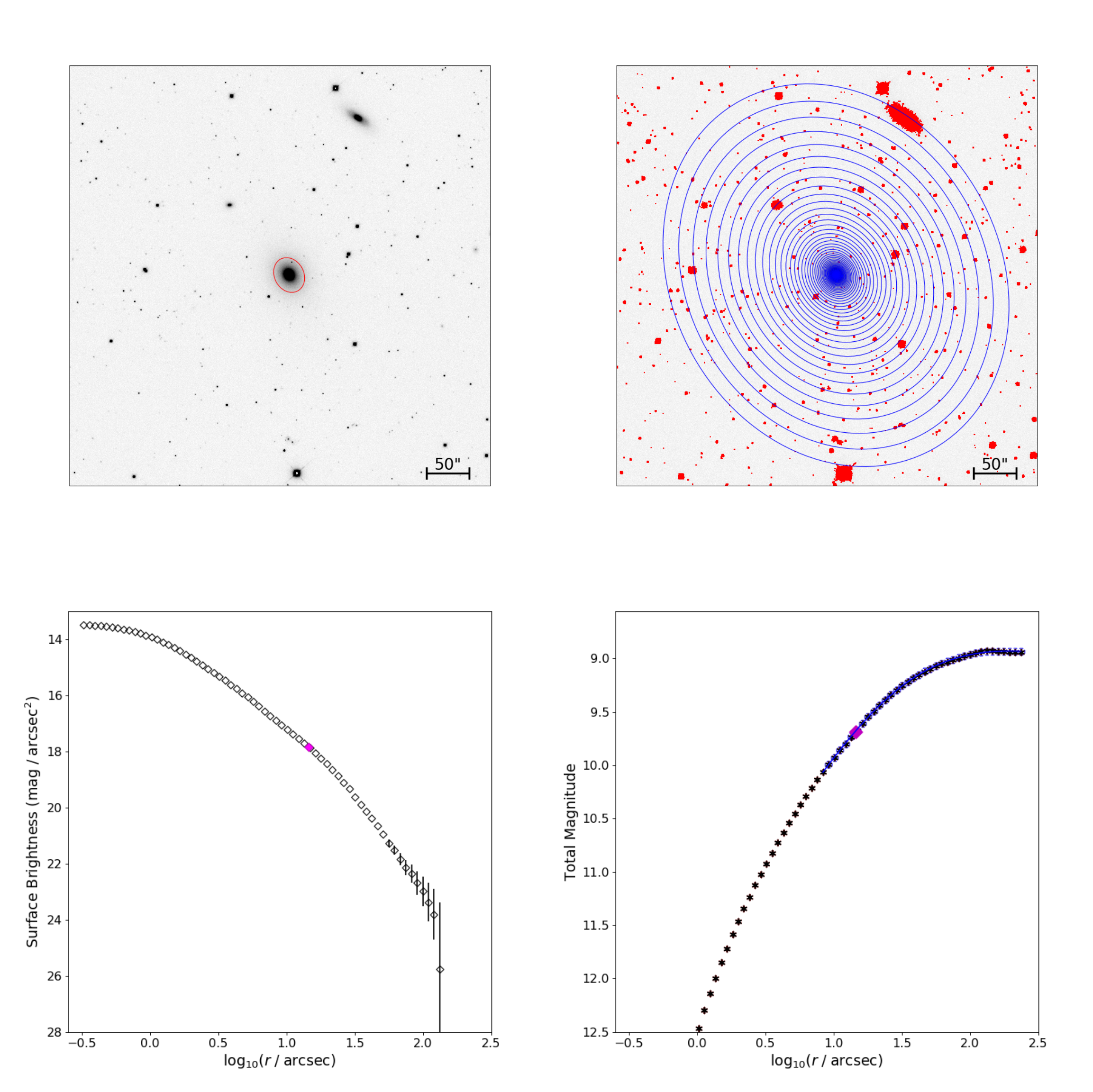}
    \caption{Example output from the photometry package ARCHANGEL for elliptical galaxy NGC~393. (Top left) The
      original CFHT WIRCam image. (Top right) Elliptical isophotes overlaid on the image, with
      masked regions in red. (Bottom left) Surface brightness profile as a function of
      isophotal semi-major axis. (Bottom right) Curve of growth, showing the enclosed magnitude as a function of isophotal semi-major axis. After iterating to improve the sky estimate, the program converges on a total $K$ magnitude for the galaxy of 8.93~mag. The black points are given by summing all pixels within a given isophote, while the blue points are given by the surface brightness interpolation described in the text. The blue line represents the rational function fit to the blue points, and the magenta diamond marks the empirically-determined half-light radius.}
    \label{fig:NGC0393}
\end{figure*}

We determine the total magnitude and half-light radius of each galaxy from the stacked image using the
galaxy photometry package ARCHANGEL \citep{Schombert2007}. Specifically, we use the
\textit{profile}, \textit{sky\_box}, \textit{bdd}, and \textit{el} routines within
ARCHANGEL.  Figure~\ref{fig:NGC0393} illustrates the basic procedure using the galaxy
NGC~393 as an example.  The steps in the procedure are discussed in more detail below.

The \textit{profile} routine produces an elliptical isophotal profile from an input
image. It begins by determining an approximate sky level in the image. Sky boxes are
scattered around the edges of the image. After discarding boxes that differ from the mean
by more than 4 standard deviations, the routine assigns the mean intensity value from these boxes as the sky intensity, and takes the standard deviation of this mean as the uncertainty on the sky value.
The routine then searches for sources in the image that exceed the
sky level by a given number of standard deviations and masks them.

Next, the routine performs the elliptical isophote fitting. The resulting ellipses are
shown in the top right panel of Figure~\ref{fig:NGC0393} for NGC~393. Ellipses are first
fit to the isophotes of the cleaned image. The ellipticities and position angles of the
ellipses are allowed to vary with radius. Any drastic changes between adjacent ellipses
are smoothed out to generate a more regular profile. Then, any pixels along each isophote
that differ from the mean intensity by more than 4 standard deviations are masked. The \textit{profile} routine
then fits a new set of ellipses to this cleaned image with the revised mask,
and the process is repeated to obtain a final cleaned image.
A final set of ellipses are then fitted to this final cleaned image and the
profile is smoothed once again. This results in an isophotal profile for the image.

In order to get an improved estimate of the sky value, we use the \textit{sky\_box} routine on the final cleaned, masked image. This routine scatters boxes that are 20 pixels by 20 pixels around the image, using the fitted isophotes to avoid the galaxy light. We find that re-fitting the sky value with the cleaned image results in a more robust determination than the preliminary fit performed in the \textit{profile} routine.

The \textit{bdd} routine takes the sky value and isophotal profile and determines the corresponding surface brightness profile. The surface brightness profile for NGC~393 is shown in the bottom left panel of Fig.~\ref{fig:NGC0393}. Parametric fits to this surface brightness profile can also be performed. 

Finally, the \textit{el} routine performs aperture photometry using the elliptical isophotes. The routine begins by filling in the masked pixels in the image with values interpolated from the isophotal profile. Then, the total luminosity within each elliptical isophote is calculated from this interpolated image. The curve of total luminosity or magnitude versus isophotal semimajor axis (shown in the bottom right panel of Fig.~\ref{fig:NGC0393}) is referred to as a curve-of-growth. Near the edge of the image, galaxy light can be dominated by sky noise. In this case, directly summing the pixels within each isophote can lead to unreliable values. To obtain a more reliable estimate, the aperture luminosities are instead estimated by interpolating the 1D surface brightness profile. This reduces noise near the outer edge of the curve of growth. 

If the galaxy light is fully contained within the field of view of the image and the sky is accurately estimated, the curve of growth will flatten at large radii. In some cases, because ARCHANGEL does not force the total luminosity to converge to a constant, the the curve of growth
does not flatten for our images. This happens in our images of 14 galaxies, with some increasing and other decreasing. Since these galaxies appear to be fully contained within the image frame, we estimate the sky from the corners of the field and adjust the adopted value within its uncertainty in order to force the curve of growth to flatten at large radii.

Once the curve of growth visually flattens, the outer end of the curve of growth is then fit with a
rational function given by a ratio of two polynomials of degree 2 as described in
\citet{Schombert2007} in order to capture the asymptotic behaviour. This fit is then evaluated at
the outermost fitted isophote to estimate the total luminosity. The semi-major axis of the
elliptical isophote containing half the total galaxy light is determined by interpolating the curve of growth. We also interpolate the surface brightness profile to determine the semi-minor axis (and thus the observed axis and ellipticity) of the half-light isophote.

\subsection{Total $K$-band Magnitudes}
\label{sec:magnitude}

We compare our total $K$-band magnitudes derived with ARCHANGEL to those from the 2MASS Extended Source
Catalog (XSC). For apparent $K$-band magnitude, we use 2MASS parameter \textit{k\_m\_ext}. This is the 
$K$-band magnitude measured within a 20 $\mathrm{mag}/\mathrm{arcsec^2}$ isophotal aperture,
extrapolated beyond the aperture via a Sersic fit.  

\begin{figure*}
	\makebox[\textwidth][c]{\includegraphics[width=1.3\textwidth]{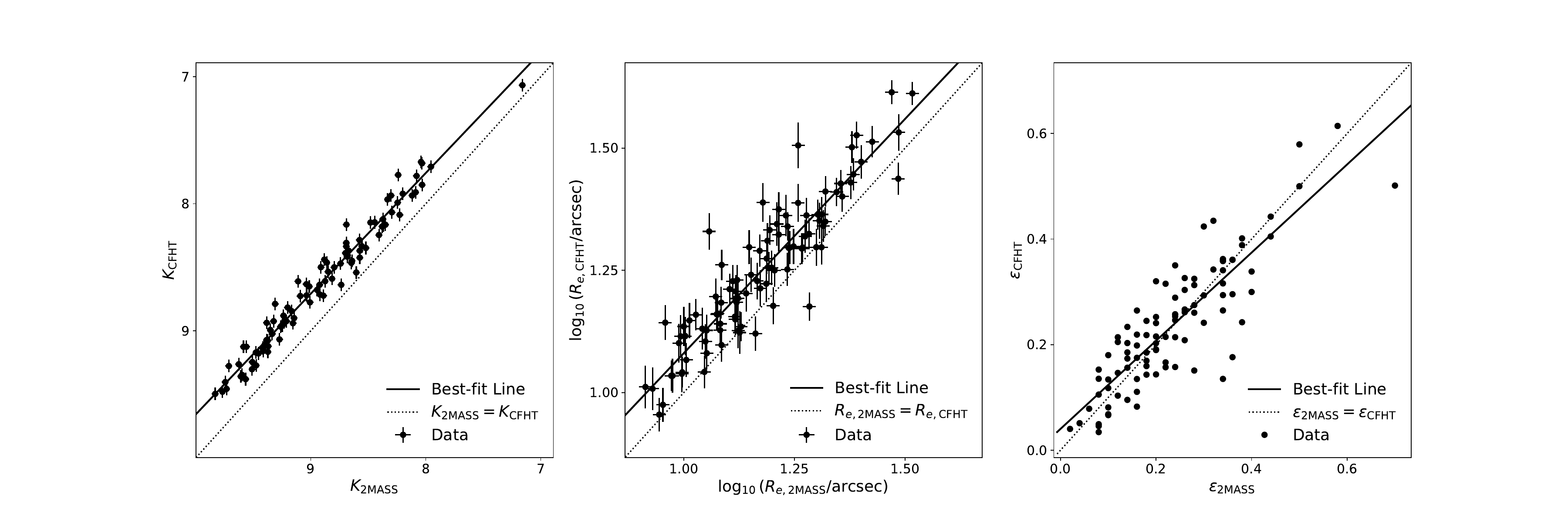}}
    \caption{Comparison of photometric properties of MASSIVE survey galaxies measured with CFHT WIRCam to 2MASS values: total $K$-band magnitudes (left), semi-major axis $K$-band half-light radii (center), and ellipticities $\varepsilon$ (right). The best-fit lines for the left and center panels are given by equations~(1) and (2), respectively. On average, $K$ from CFHT is brighter than 2MASS by $0.29\pm0.01$ mag, $R_e$ from CFHT is 18\% larger than 2MASS, and no significant offset is observed between CFTH and 2MASS for ellipticity.}
    \label{fig:2mass}
\end{figure*}

The galaxy magnitudes measured from the CFHT curve of growth are systematically brighter than the corresponding magnitudes from the 2MASS survey. This is demonstrated in the left panel of Fig.~\ref{fig:2mass}. The best-fit linear relation between the apparent $K$-band magnitudes from CFHT and 2MASS is given by
\begin{equation}
(K^\mathrm{CFHT}-9) = b_{K}(K^\mathrm{2MASS}-9) + a_{K} \,,
\end{equation}
where the best-fit values of these parameters are $a_K=-0.292\pm0.009$ and $b_K=0.950\pm0.017$ as determined using the linear fitting procedure outlined in Section~\ref{sec:fitting}. The slope is mildly inconsistent with unity. 
$K^\mathrm{CFHT}$ has an average scatter of 0.09 about this relation. On average, $K^\mathrm{CFHT}$ is brighter than $K^\mathrm{2MASS}$ by $0.29\pm0.01$ mag. 

\citet{SchombertSmith2012} report a bias in the $J$-band total magnitudes extracted from the 2MASS XSC. The
surface brightness profiles extracted from 2MASS are systematically too dim, particularly at larger
radii. The source of this discrepancy is determined to be an issue with the 2MASS sky subtraction
scheme. Consequently, total $J$ magnitudes from 2MASS are found to be systematically too faint by an average of 0.33~mag.  These authors do not provide a direct comparison for the $K$ band, but their discussion suggests that
the size of the offset should be similar.

The offset that we measure at $K_\mathrm{2MASS}\sim10$ mag is $0.34{\,\pm\,}0.02$ mag, consistent
the $J$ band offset reported in \citet{SchombertSmith2012}.
However, we find the slope of the relation between $K_\mathrm{CFHT}$
and $K_\mathrm{2MASS}$ to be slightly smaller than 1.0 (equation~1), meaning the offset between the two sets of magnitudes decreases
for the brightest galaxies in our sample. For example, $K_\mathrm{CFHT}-K_\mathrm{2MASS}\sim -0.2$
mag at $K_\mathrm{2MASS}\approx7.5$ mag. For giant ellipticals, this corresponds to 
$J_\mathrm{2MASS}\approx8.5$ mag.  Examination of Fig.~12 from \citet{SchombertSmith2012}
reveals that the offset for the brightest $\sim\,$2~mag of their comparison with 2MASS
is significantly smaller than the quoted mean and is approximately
$J-J_\mathrm{2MASS}\sim-0.2$ mag, which is very similar to what we find in our $K$-band comparison.

Comparable 2MASS magnitude offsets have been observed in other samples as
well. \citet{RiosLopezetal2021} use Galfit \citep{Pengetal2002} to measure total extrapolated magnitudes from 2MASS
images for 101 bright, nearby galaxies, including 20 early-type galaxies. They find no significant
offset with respect to the 2MASS XSC total magnitudes for late-type galaxies, but they do find
significant offsets for the smaller sample of early-type galaxies.  In the $K$-band, they report a
mean offset of $0.34\pm0.07$ mag; however, the scatter is large, and the sample includes several
intrinsically faint nearby galaxies, including two Local Group dwarf ellipticals.  If we limit the
comparison to the subsample of 17 early-type galaxies at distances $\gtrsim10$~Mpc to make it more similar to
our sample, then the weighted offset is $0.36{\,\pm\,}0.08$ mag, with a scatter of 0.32 mag and a
median offset of 0.27~mag.  The scatter is several times larger than we find, but the offset agrees well
with our result.

\citet{Lasker2014}, again using Galfit to do 2-D parametric modeling, measure extrapolated $K$-band
magnitudes from CFHT/WIRCam data for a sample of 35 nearby galaxies of all morphological types with
well-measured central black hole masses.  These authors report several versions of the total
magnitudes, including magnitudes derived from a single S\'ersic model, a ``standard bulge plus
disk'' model, and an ``improved'' model that adds structural components (up to six for galaxies with complex structure)
to the standard model until the residuals over the field are judged visually to be at an
acceptable level.  They quote an average offset of 0.34 mag for the total magnitudes from
their ``improved'' models with respect to 2MASS, in the sense that 2MASS is too faint.

If we omit the spirals and consider only the 31 galaxies that \citet{Lasker2014} classify as either
elliptical or S0, then the mean offset (unweighted, as the authors do not provide uncertainties)
with respect to the magnitudes from their ``improved'' models is 0.34~mag, the median offset is 0.25~mag,
and the scatter is 0.28~mag.  However, we note that for this subset of early-type galaxies,
the standard bulge+disk models in \citet{Lasker2014} give better agreement with the 2MASS total
magnitudes, with a scatter of 0.20~mag, or 40\% less than the scatter given by their
``improved models'' with the additional structural components.
Using the set of standard model magnitudes, the mean offset for the 31
early-types is $0.27\pm0.04$ mag, and the median offset is also 0.27~mag.  This is similar to the
mean offset that we find with respect to 2MASS, although the scatter in our comparison is a factor of
two lower.

Four of the galaxies in the MASSIVE survey are studied in \citet{Lasker2014}: NGC~4486 (M87), NGC~4649 (M60), NGC~5252, and NGC~7052. Of these, we obtained new CFHT data for NGC~5252 and NGC~7052.
For NGC~5252, our $K$-band total
magnitude is 0.29~mag brighter than the result from their standard model, and 0.11~mag fainter than
their ``improved'' model (which is 0.40 mag brighter than 2MASS).  For NGC~7052, our measurement is
0.07~mag fainter than their standard model, and they do not make any improvements to
that model.  Thus, from this very small direct comparison, the offsets are consistent with the
scatter.

We conclude that our measured offset of $0.29\pm0.01$~mag with respect to the $K$-band total
magnitude from the 2MASS XSC agrees well with previous studies. The scatter we find of 0.09~mag is
considerably less than previous comparisons using parametric modeling.  Finally, for the subset of early-type galaxies in \citet{Lasker2014}, the standard bulge+disk models appear to give more robust magnitudes than the ``improved'' models.

\subsection{Half-light Radii}
\label{sec:Re}
The half-light radii derived from CFHT can be compared to those of the 2MASS XSC. The 2MASS catalog lists semi-major axes of the half-light elliptical isophote in 3 different bands (listed as
parameters j\_r\_eff, h\_r\_eff, and k\_r\_eff). We take the $K$-band half-light radius, k\_r\_eff,
to compare with the semi-major axis $R_e$ determined from our CFHT $K$-band photometry.

We find $R_e$ from CFHT to be systematically larger than from 2MASS (center panel of Fig.~\ref{fig:2mass}). The best-fit power-law relation between the 2MASS and CFHT half-light radii is given by:
\begin{equation}
\log_{10}{\left(\frac{R_e^\mathrm{CFHT}}{10^{1.2}\;\mathrm{arcsec}}\right)} = b_{R_e}\log_{10}{\left(\frac{R_e^\mathrm{2MASS}}{10^{1.2}\;\mathrm{arcsec}}\right)} +a_{R_e}
\end{equation}
where the best-fit values of these parameters are $a_{R_e}=0.072\pm0.006$ and
$b_{R_e}=0.96\pm0.04$. The slope is consistent with unity to within $1\sigma$. $R_e^\mathrm{CFHT}$
has a mean scatter of 0.06 dex about this relation. On average, 
$R_e^\mathrm{CFHT}$
is a factor of $10^{0.072} = 1.18$ that of $R_e^\mathrm{2MASS}$. Since half-light radii are
derived from total magnitudes,
the larger $R_e$ from CFHT is correlated with the brighter CFHT $K$ in Sec.~\ref{sec:magnitude}.

In some contexts, it is useful to consider the geometric, or circularized radius of the half-light
isophote. The ellipticity of the half-light isophote is needed to determine this from the major
axis. We compare the elliptical axis ratios of our isophotes to those reported by 2MASS. For this,
we use the sup\_ba parameter from the 2MASS catalog. Our average observed flattening is in
excellent agreement with 2MASS. There is no significant shift between the two. The CFHT values
display a scatter of 0.06 about the 2MASS values. 

\input{tables/main_table.tex}

\subsection{Conversion to Physical Parameters}

The main galaxy parameters of interest to us in this paper are the total absolute $K$-band magnitude $M_K$ (or, equivalently, the $K$-band luminosity $L_K$) and the
semi-major axis half-light radius $R_e$ (or, alternatively, the circularized version
$R_{e,\mathrm{circ}}$). To convert our measurements to physical physical properties 
requires a set of reliable distances.
We use distances from \citet{Jensenetal2021}, where available. These distances are derived using
surface brightness fluctuations (SBF) in WFC3 Hubble Space Telescope images, and have a median
uncertainty of $3.9\%$. For galaxies not studied in \citet{Jensenetal2021}, we use the values
reported in the first MASSIVE paper \citep{Maetal2014}. In \citet{Maetal2014}, the SBF distances are used when available;
otherwise, distances from group-corrected flow
velocities are used. In the following, we adopt a $4\%$ uncertainty on SBF derived distances, and a $10\%$ uncertainty on those derived from group-corrected flow velocities. 

Table~\ref{tab:param_vals} lists our photometric and structural measurements, the
physical parameters derived from these measurements, as well
as other properties of the MASSIVE survey galaxies.  The following sections discuss the 
relations between these parameters. For conversion from $M_K$ to $L_K$ in solar units, we use $M_{K,\odot} = 3.28$~mag \citep[e.g.,][]{Willmer2018} with the standard definition: $L_K = 10^{-0.4(M_K-M_{K,\odot})} L_{K,\odot}$.

\section{Determining Galaxy Scaling Relations}
\label{sec:scalingrelations}

\subsection{Uncertainties}

In order to study the relationships among the photometric parameters determined in the previous Section and other galaxy properties, we need to estimate the uncertainties on these measurements. For most of our galaxy images, the dominant source of uncertainty is sky estimation. By comparing sky estimates from the automated \textit{sky\_box} routine, manually placed sky boxes and the asymptotic intensity of the isophotal profile, we find that our sky estimates result in values of $K$ with an uncertainty of about $\delta K = 0.05$ mag. The uncertainty in the half-light radius is also dominated by the uncertainty in the sky values. We thus assume the uncertainties in these two measurements to be perfectly correlated. We measure the slope of the curve-of-growth (aperture magnitude versus logarithm of aperture semi-major axis) at $\log{(\Rapp)}$, $S=\left|\frac{\mathrm{d}K(R)}{\mathrm{d}\log{(R)}}\right|_{\log{(\Rapp)}}$ and assign a corresponding uncertainty on $\log{(\Rapp)}$ of $\delta \log{(\Rapp)} = S \delta K$. This results in a median uncertainty on $\Rapp$ of about $8\%$. We therefore adopt a covariance matrix on these quantities given by
\begin{equation}
\label{eq:apparent_covariance}
    \mathrm{Cov}(\log{(\Rapp)},K)=\begin{pmatrix}
    (S\delta K)^2 & -S(\delta K)^2 \\
    -S(\delta K)^2 & (\delta K)^2\,, \\
    \end{pmatrix}
\end{equation}
where the corresponding absolute quantities, $R_e$ and $M_K$ or $L_K$, will also have correlated uncertainties. This covariance arises for two reasons: the measurements of the apparent magnitude $K$ and angular size $\Rapp$ are strongly correlated, and the intrinsic quantities $R_e$ and $M_K$ both depend on the distance. The total covariance matrix between these two quantities is given by:
\begin{equation}
\begin{split}
    \mathrm{Cov}&(\log{(R_e)},M_K)\\
    &=\mathrm{Cov}(\log{(\Rapp)},K)+(\delta \log{D})^2\begin{pmatrix}
    1 & -5 \\
    -5 & 25 \end{pmatrix} \\
    &=\begin{pmatrix}
    (S\delta K)^2+(\delta \log{D})^2 & -S(\delta K)^2-5(\delta \log{D})^2 \\
    -S(\delta K)^2-5(\delta \log{D})^2 & (\delta K)^2+25(\delta \log{D})^2 \\
    \end{pmatrix} \,.
\end{split}
\end{equation}
For our given uncertainties, this results in a mean correlation coefficient between the uncertainties on $\log{(R_e)}$ and $M_K$ of about 0.87. Ignoring the correlation between the apparent quantities and only including the correlation due to the distance estimation would still result in a median correlation of 0.39 for galaxies with distances measured via SBF and 0.76 for galaxies with distances measured via group-corrected flow velocities.

\subsection{Linear Fitting Procedure}
\label{sec:fitting}

Since our data can have correlated uncertainties in both the dependent and independent variables, as well as a selection on absolute magnitude, $M_K$, a robust fitting procedure is needed in order to obtain reliable results.

The fitting procedure we use throughout this paper is the \textit{LinMix} procedure outlined in \citet{Kelly2007}. This procedure constructs a likelihood function for the data, in which the distribution of the independent variable is modelled as a mixture of gaussian functions. The dependent variable is then assumed to be drawn from a gaussian distribution centered on a linear relation with respect to the independent variable. The main strengths of this procedure are its explicit model for the data, and its ability to account for both selection effects and covariances between the uncertainties on the dependent and independent variables. It also returns samples from the posterior distribution over parameters, allowing for a clearer interpretation of the fit uncertainties. For each parameter, we report a marginalized 1D best-fit value and uncertainty. The best-fit value is determined as the 50th percentile of the 1D distribution of samples, while the 1$\sigma$ uncertainties are determined from the 16th and 84th percentiles.

\section{Size-Luminosity Relation}
\label{sec:SL}

\begin{figure*}
\includegraphics[width=\textwidth]{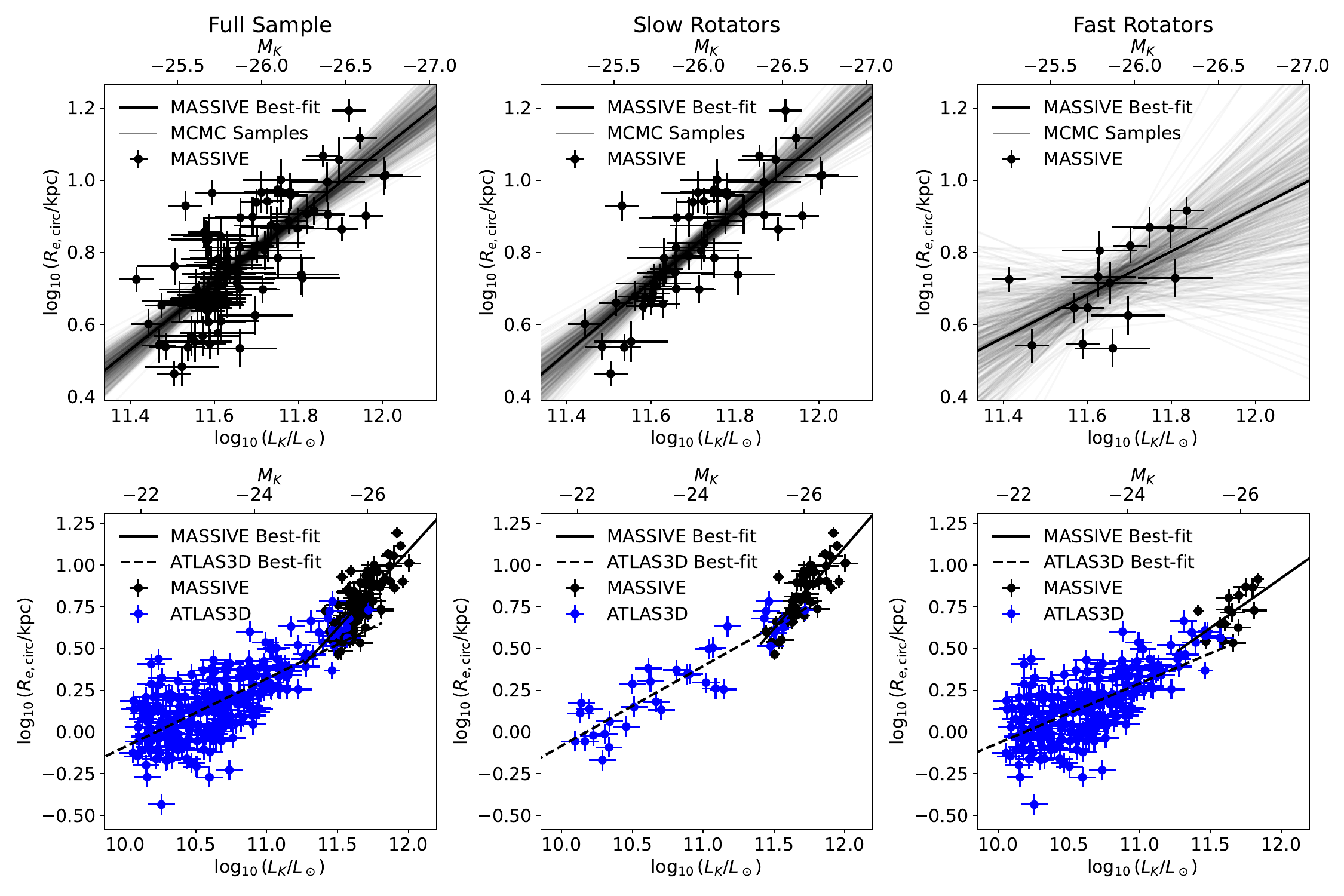}
\caption{Size-luminosity relation for early-type galaxies in the MASSIVE survey (top) and in the MASSIVE and \atlas surveys together (bottom; black for MASSIVE, blue for \atlas). The full samples (left column) are separated into slow rotators (middle) and fast rotators (right). The best-fit linear relations to various samples are summarized in
Table~\ref{tab:SL_params}. In the top row, the grey lines represent MCMC draws from the posterior distribution over the parameters describing the linear relations for MASSIVE galaxies. The size-luminosity relation shows a clear steepening at $L_K \sim 10^{11.5} L_{K,\odot}$ in the full sample as well as in the subsample of slow rotators. Lower-luminosity galaxies within the \atlas survey are oversized compared to the MASSIVE linear relations.
    }
    \label{fig:SL}
\end{figure*}

We consider a power-law size-luminosity (SL) relation of the form
\begin{equation}
\label{eq:SL}
\log_{10} {\left(\frac{R_{e,\mathrm{circ}}}{1\,\mathrm{kpc}} \right)} = b_\mathrm{SL}\log_{10} {\left( \frac{L_K}{10^{11.5}L_{K,\odot}}\right)} + a_\mathrm{SL}  \,,
\end{equation}
where $R_{e,\mathrm{circ}}$ is the geometric radius of the half-light ellipse and
$L_K$ is the total $K$-band luminosity of a galaxy.
We further assume an intrinsic normal scatter in $\log_{10}{(R_{e,\mathrm{circ}})}$ about this relation with standard deviation $\epsilon_\mathrm{SL}$.

\subsection{MASSIVE and \atlas Galaxies}
\label{sec:SL_fits}

\begin{table}
  \begin{tabular}{|l|lll|}
    \hline
    Sample & $a_\mathrm{SL}$ & $b_\mathrm{SL}$ & $\epsilon_\mathrm{SL}$ \\ \hline
    MASSIVE (full) & $0.623^{+0.019}_{-0.018}$ & $0.93\pm0.10$ & $0.099^{+0.008}_{-0.007}$ \\
    \atlas (full) & $0.52\pm0.02$ & $0.41\pm0.03$ & $0.143^{+0.007}_{-0.006}$ \\ \hline
    MASSIVE (SRs) & $0.62\pm0.03$ & $0.98\pm0.11$ & $0.096^{+0.012}_{-0.010}$ \\
    \atlas (SRs) & $0.63\pm0.03$ & $0.48\pm0.04$ & $0.115^{+0.017}_{-0.014}$ \\ \hline
    MASSIVE (FRs) & $0.62^{+0.06}_{-0.05}$ & $0.59\pm0.3$ & $0.11^{+0.03}_{-0.02}$ \\
    \atlas (FRs) & $0.47\pm0.03$ & $0.36\pm0.03$ & $0.144^{+0.008}_{-0.007}$ \\ \hline
    \end{tabular}
    \caption{Fit parameters for the size-luminosity relation, as defined by equation~(\ref{eq:SL}), in various galaxy samples. The first and second rows represent the MASSIVE and \atlas surveys, respectively. The third and fourth rows represent the SRs in the MASSIVE and \atlas surveys, and the fifth and sixth rows represent the FRs in the two surveys. The FR and SR classifications are described in the text.}
    \label{tab:SL_params} 
\end{table}

Fig.~\ref{fig:SL} (top left panel) shows  $R_{e,\mathrm{circ}}$ versus $L_K$ for MASSIVE galaxies and our best-fit SL relation; linear relations corresponding to a number of samples from the posterior distribution are shown by the grey lines distributed about the best-fit relation. The best-fit parameter values and associated uncertainties for the SL relation for the full MASSIVE sample are listed in row 1 of Table~\ref{tab:SL_params}. The uncertainties in $a_\mathrm{SL}$ and $b_\mathrm{SL}$ are strongly correlated. 

One of the defining selection criteria for the MASSIVE survey is the absolute $K$-band magnitude
selection, $M_K<-25.3$ mag, as measured in the 2MASS XSC. 
To extend this dynamic range, we compare with galaxies from the \atlas survey, a volume-limited sample of early-type galaxies with $M_K<-21.5$. MASSIVE includes galaxies out to a distance of $\sim 100$ Mpc, while \atlas includes galaxies out to
42~Mpc. The similarity in selection criteria makes \atlas a natural comparison sample to MASSIVE.  

A fair comparison of the relations for MASSIVE and \atlas galaxies requires
consistent sets of measurements for the two surveys. For the half-light radii of \atlas\ galaxies, we use the values reported by \citet{Cappellari2011}, which are taken from the RC3 \citep{deVaucouleursetal1991} where available, and from 2MASS otherwise. The RC3 values
are based primarily on $B$-band photoelectric photometry extrapolated to infinite aperture and then
interpolated to find the equivalent circular aperture containing half the total light, without
corrections for nearby contaminating sources.  \citet{Cappellari2011} find that the RC3 values are a factor of 1.7 times larger than 2MASS on average (with a scatter of 0.11~dex) and scale the values from 2MASS for their galaxies up by a factor of 1.7. In order to directly compare their radii for \atlas\ galaxies with our radii for MASSIVE galaxies as measured with CFHT, we divide their radii by a factor of 1.7 and then multiply by a factor of 1.18 (see Section~\ref{sec:Re}) to agree with CFHT on average. Note that this procedure assumes that the same scaling relation between 2MASS and our CFHT radii holds for lower luminosity galaxies in \atlas\ that appear similar in angular size to our sample because they are more nearby.

The $K$-band luminosities of \atlas\ galaxies are taken from 2MASS without corrections \citep{Cappellari2011}. 
We apply a constant shift of 0.29 mag to these values such that they agree with the average offset between 2MASS and our CFHT magnitudes. 

The resulting $R_e$ and $L_K$ for \atlas\ galaxies are added to the MASSIVE sample in the bottom left panel of Figure~\ref{fig:SL}.
Brighter than $M_K\sim -25$ mag, the two samples appear to agree well with no visually obvious offset or change in slope. Below $M_K\sim -24$ mag, the \atlas galaxies exhibit a clear change in slope. 
% This is qualitatively similar to the non-linear zone-of-exclusion observed in the size-mass plane in \citet{Cappellarietal2013b} using dynamically determined total stellar masses.
Despite apparent non-linearity in \atlas at low luminosities, we report the best-fitting linear relation parameters in row 2 of Table \ref{tab:SL_params} for completeness.

\subsection{Fast vs. Slow Rotators}

We now examine the SL relation for the slow rotators (SRs) and fast rotators (FRs) separately. 
The \atlas galaxies are classified as FRs and SRs in \citet{Emsellemetal2011}. The
classification scheme for MASSIVE galaxies is described in \citet{Vealeetal2017b}.
The two differ only in minor differences in the apertures used to measure the rotation. 

The MASSIVE sample contains 17 fast rotators and 59 slow rotators that also have CFHT $K$-band photometric data.  The \atlas sample has 224 fast rotators and 36 slow rotators.

The center two panels of Fig.~\ref{fig:SL} show the SL relation for SRs in the MASSIVE survey (top panel) and in the combined MASSIVE and \atlas\ sample (bottom panel).
The slope of the SL relation for the MASSIVE SRs alone is consistent with unity:
$b_{\rm SL} = 0.98\pm 0.11$ (row 3 of Table~\ref{tab:SL_params}). This slope
is consistent with the slope of $0.93\pm 0.10$ for the full MASSIVE sample.
The \atlas SRs, on the other hand, follow a much shallower SL relation with a slope of
$0.48\pm 0.04$ (row 4 of Table~\ref{tab:SL_params}). Since \atlas\ and MASSIVE galaxies cover (nearly) distinct ranges of $L_K$, there is a significant difference in the slope of the SL relation between low luminosity (\atlas) and high luminosity (MASSIVE) galaxies. 

Due to the small number of FRs within the MASSIVE sample (top right panel of Fig.~\ref{fig:SL}), the slope of the SL relation for FRs alone is quite uncertain: $b_{SL}=0.59\pm0.3$ (row 5 of Table~\ref{tab:SL_params}).
FRs in both MASSIVE and \atlas 
samples are shown in the bottom right panel of Fig.~\ref{fig:SL}.
The best-fit linear SL relation for the \atlas FRs has a slope of $0.36\pm 0.03$
(row 6 of Table~\ref{tab:SL_params}), which 
is within $1\sigma$ uncertainty of the slope
for the MASSIVE FRs.
Visual inspection suggests a change in slope around $L_K\sim10^{10.75}L_{K,\odot}$, with MASSIVE and \atlas galaxies lying on the same relation above this break.

\subsection{Discussion}

Several studies have examined the local SL relation for galaxies in the Sloan Digital Sky Survey (SDSS).  In the $r$-band,
the SDSS SL relation is found
to be $R_e\propto L^{0.68}$ for
elliptical galaxies as a whole, and $R_e\propto L^{0.88}$ for BCGs alone
\citep{Bernardietal2003, Bernardietal2007}.
More recently, \citet{Bernardietal2014} analyzed 
the bulge components of early-type galaxies in SDSS and 
found $R_e\propto L_\mathrm{bulge}^{0.85}\,$, with a comparable scaling in total luminosity among the most luminous early type galaxies.

The nearly linear SL relation (in $K$-band) that we find for MASSIVE galaxies (either full sample or SRs) is consistent with the steeper relation from SDSS for BCGs and the bulge components of a broader range of early-type galaxies, but it is significantly steeper than the $r$-band relation for early-type galaxies as a whole within SDSS. \citet{Bernardietal2007} suggest that this curvature may arise from an increased fraction of BCGs at high-luminosity, together with the fact that BCGs tend to have larger sizes than expected from global scaling relations for ellipticals. Along these lines, deep imaging of some nearby BCGs, find significantly larger $R_e$ values then estimated from previous shallow surveys \citep[e.g.][]{Iodice2016,Spavone2017}. This may be due to the presence of diffuse, extended halos that change the shape of the surface brightness profiles at large radius.

To assess whether BCGs in the MASSIVE survey exhibit a distinct scaling relation from other ellipticals, we adopt the classification scheme of \citet{Vealeetal2017a}, in which the MASSIVE galaxies are classified as Brightest Group Galaxies, isolated galaxies, or satellite galaxies based on the high density contrast (HDC) group catalogue \citep{Crooketal2007}. These classifications are based on whether galaxies belong to a group of three or more members (with $K<11.25$), and whether the galaxy is the most luminous in this group. Within this classification scheme, only 21\% of galaxies are classified as satellites. Due to the small number of satellite galaxies in the MASSIVE sample, our fits are not able to determine whether BCGs exhibit a different scaling relation than satellite ellipticals.

A steepening in slope for the SL relation as a function of luminosity has been reported in a number of other studies. For example, \citet{Lauer2007a} finds a power-law slope of $b_\mathrm{SL}=(0.50\pm0.08)$ at fainter $V$-band magnitudes and a much steeper slope of $b_\mathrm{SL}=(1.18\pm0.06)$ for more luminous ellipticals. The transition between the two slopes occurred at $M_V \approx -22$~mag, corresponding to $M_K \approx -25.3$~mag at the typical color of giant ellipticals \citep[e.g.,][]{Michard2005}. Similar curvature has been suggested in the $K$-band using the shallow 2MASS photometry \citep{Forbesetal2008}.
Curvature in the SL relation was also reported for early-type galaxies in the SDSS by \citet{HydeBernardi2009}. 
\citet{GrahamGuzman2003} and \citet{GrahamWorley2008} show that a curved SL relation can
result from two parameters having single power-law relationships with luminosity: the central surface
brightness of an inward interpolation of the best-fit outer Sersic profile, and the Sersic index $n$. 
Testing this hypothesis in the current context would require fitting core-Sersic models to the surface brightness profiles of the combined MASSIVE+\atlas\ galaxy sample.

\section{Faber-Jackson Relation}
\label{sec:FJ}

\begin{figure*}
\includegraphics[width=\textwidth]{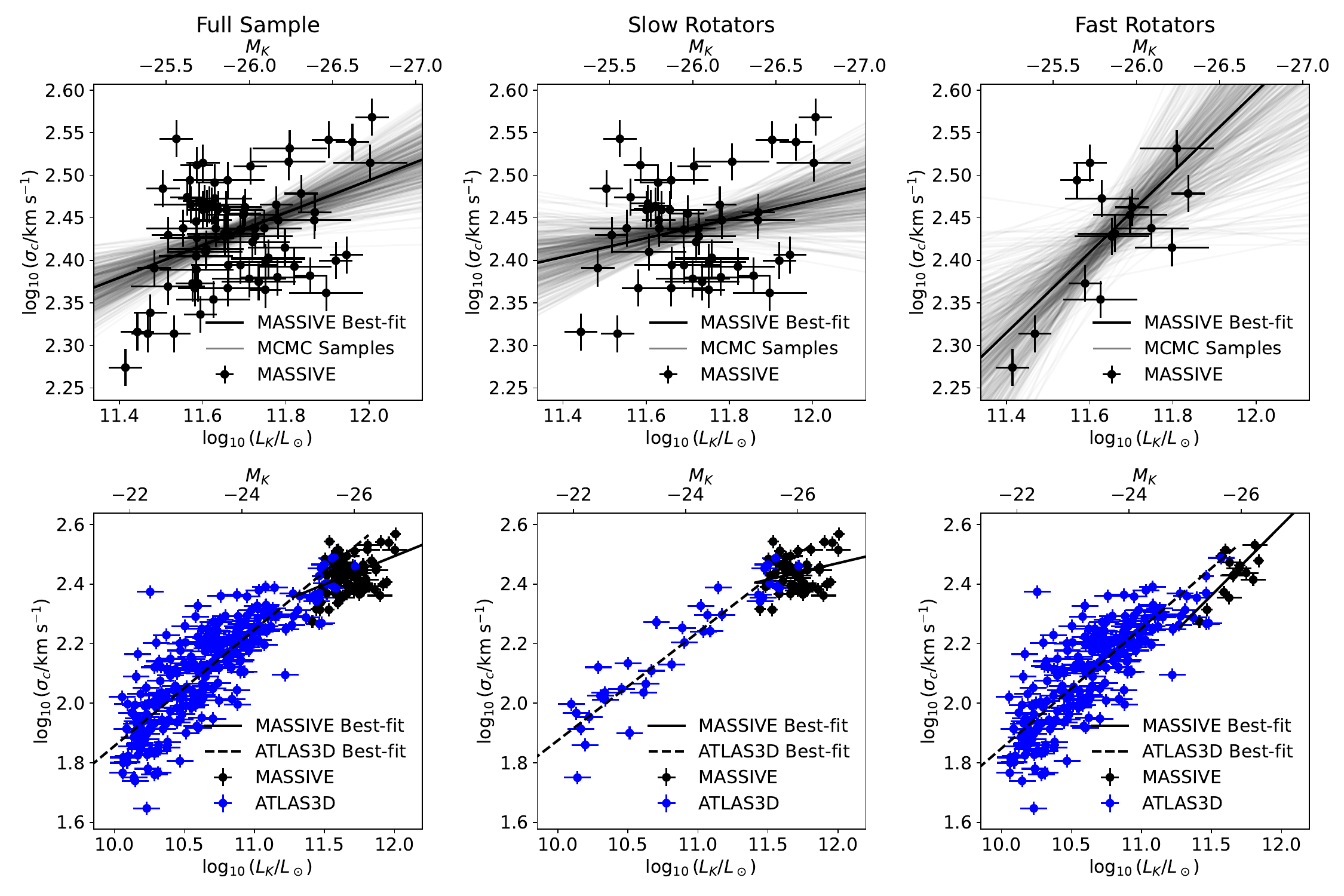}
\caption{
Faber-Jackson relation for early-type galaxies in the MASSIVE survey (top) and the MASSIVE and \atlas surveys together (bottom; black for MASSIVE, blue for \atlas). 
The full samples (left column) are separated into slow rotators (middle) and fast rotators (right). The best-fit linear relations to various samples are summarized in Table~\ref{tab:FJ_params}. In the top row, the grey lines represent MCMC draws from the posterior distribution over the parameters describing the linear relations for MASSIVE galaxies.
Lower-luminosity galaxies within the \atlas survey have lower velocity dispersions than would be predicted by the linear relations for the full and SR MASSIVE samples.}
\label{fig:FJ}
\end{figure*}

\begin{table}
    \begin{tabular}{llll}
    \hline
    Sample & $a_\mathrm{FJ}$ & $b_\mathrm{FJ}$ & $\epsilon_\mathrm{FJ}$ \\ \hline
    MASSIVE (full) & $2.399\pm0.012$ & $0.19\pm0.06$ & $0.051\pm0.005$ \\
    \atlas\ (full) & $2.445\pm0.016$ & $0.39\pm0.02$ & $0.090\pm0.005$ \\ \hline
    MASSIVE (SRs) & $2.415\pm0.016$ & $0.11\pm0.07$ & $0.054^{+0.007}_{-0.006}$ \\
    \atlas (SRs) & $2.423\pm0.02$ & $0.36\pm0.02$ & $0.063^{+0.012}_{-0.011}$ \\ \hline
    MASSIVE (FRs) & $2.36\pm0.03$ & $0.47\pm0.18$ & $0.056^{+0.019}_{-0.014}$ \\
    \atlas (FRs) & $2.46\pm0.02$ & $0.40\pm0.02$ & $0.093^{+0.006}_{-0.005}$ \\ \hline
    \end{tabular}
    \caption{Fit parameters for the Faber-Jackson relation, as defined by equation~(\ref{eq:FJ}).  The galaxy subsamples are the same as in Table~\ref{tab:SL_params}.}
    \label{tab:FJ_params}
\end{table}

We consider a Faber-Jackson relation of the form
\begin{equation}
\label{eq:FJ}
\log_{10} \left({\frac{\sigma_c}{1\mathrm{km\;s^{-1}}}} \right) = b_\mathrm{FJ}\log_{10} \left({\frac{L_K}{10^{11.5}L_{K,\odot}}} \right) + a_\mathrm{FJ} \,,
\end{equation}
where $\sigma_c$ is the central velocity dispersion and $L_K$ is the total $K$-band luminosity of a galaxy.  We further assume an intrinsic normal scatter in $\log_{10}{(\sigma_c)}$ about this relation with standard deviation $\epsilon_\mathrm{FJ}$.

Similar to the previous section, we consider both MASSIVE and \atlas galaxies. For the velocity dispersions of the MASSIVE galaxies, we use the values reported in \citet{Vealeetal2018}. Mainly, we use the velocity dispersion measured within the central fiber of the Mitchell IFU, $\sigma_c$. We also use $\sigma_e$, which is measured via a luminosity-weighted average of $\sigma$ for fibers within a radius $R_{e,\mathrm{NSA}}$ of the galaxy center\footnote{Note that this luminosity-weighted average is not the same as determining the velocity dispersion for a single spectrum with radius $R_{e,\mathrm{NSA}}$. For slow-rotating galaxies, the difference is very minor. For fast-rotating galaxies, the difference can be significant.}. Here, $R_{e,\mathrm{NSA}}$ is the half-light radius reported by the NASA-Sloan Atlas (NSA), based on the SDSS DR8 catalogue \citep{Aiharaetal2011}. Where values from NSA were not available, values from 2MASS were used and corrected for the relative slope and offset between the two using equation~4 of \citet{Maetal2014}. We do not attempt to correct these values to the newly determined radii reported above. In the following, we adopt a 5\% uncertainty on both $\sigma_c$ and $\sigma_e$ values.

For \atlas galaxies, we adopt the velocity dispersion measured within a circular aperture of radius 1 kpc \citep{Cappellarietal2013b}. This aperture is only moderately larger than the average aperture used for the MASSIVE galaxies which is typically between about 0.6 kpc and 1 kpc depending on the distance to the galaxy.

\subsection{MASSIVE and \atlas Galaxies}

The FJ relation for the MASSIVE sample and the combined sample is shown in Fig.~\ref{fig:FJ}. For MASSIVE galaxies alone, we find $\sigma_c \propto L^{0.19\pm 0.06}$ (row 1 of Table~3). 
The brightest \atlas galaxies (above $M_K\sim-24$ mag) follow this relation closely.
However, there is a prominent curvature at lower luminosities, with lower luminosity \atlas galaxies having smaller central velocity dispersion than predicted from extrapolation of the linear relation from MASSIVE. 
A single power-law fit to the \atlas sample gives a significantly different slope, $\sigma_c \propto L^{0.39\pm 0.02}$ (row 2 of Table~3). Since the FJ relation within the \atlas sample shows a curvature, this slope should be treated as an average value for \atlas.
We note that a change in the slope of the mass versus velocity dispersion relation was earlier reported by \citet{Cappellarietal2013b}, and thus our result does not simply indicate a nonlinear behavior in $M/L$.

\begin{figure}
	\includegraphics[width=0.9\linewidth]{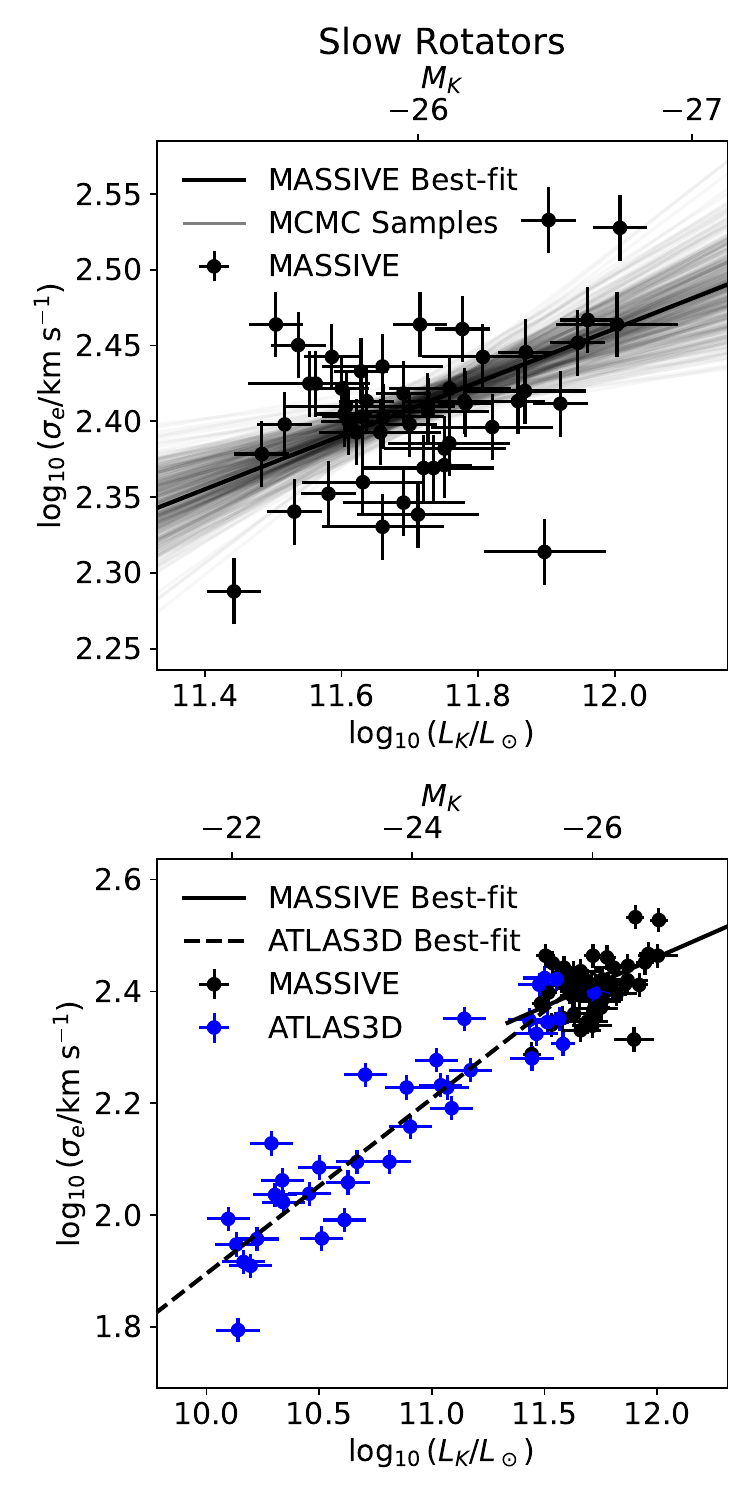}
    \caption{Same as the FJ relation for slow rotators in Fig.~\ref{fig:FJ}, 
    except the central velocity dispersion $\sigma_c$ is replaced with $\sigma_e$ measured within the half-light radius. 
    The best-fit line for MASSIVE is given by the parameters $a_\mathrm{FJ}=2.373\pm0.012$, $b_\mathrm{FJ}=0.18\pm0.05$, and
    $\epsilon_\mathrm{FJ}=0.037^{+0.006}_{-0.005}$. Low luminosity
    galaxies within the \atlas survey have lower dispersions than would be predicted by an
    extrapolation of the linear relation from MASSIVE. The best-fit line for \atlas is given by the
    parameters  $a_\mathrm{FJ}=2.368^{+0.011}_{-0.009}$, $b_\mathrm{FJ}=0.31\pm0.02$, and
    $\epsilon_\mathrm{FJ}=0.057^{+0.011}_{-0.009}$.} 
    \label{fig:FJ_e_slow}
\end{figure}

\subsection{Fast vs. Slow Rotators}

We now examine the FJ relation for the slow and fast rotators separately. These subsamples are also shown in Fig.~\ref{fig:FJ}.
For SRs, there is a significant change in slope between the MASSIVE and \atlas samples, with $\sigma_c$ increasing less quickly at larger luminosities. This trend is similar to that seen for the full samples in the previous subsection.  The MASSIVE SRs follow 
$\sigma_c \propto L^{0.11\pm 0.07}$ (row 3 of Table~3), which is consistent with, but moderately shallower than, the slope for the full MASSIVE sample.  The \atlas SRs follow 
a different relation from MASSIVE SRs: $\sigma_c \propto L^{0.36\pm 0.02}$ (row 4 of Table~3).   

Due to the small number of FRs in the MASSIVE sample, the slope for MASSIVE FRs has large uncertainties: $\sigma_c \propto L^{0.47\pm 0.18}$ (row 5 of Table~3).  In comparison, a single power-law fit to the \atlas FRs gives $\sigma_c \propto L^{0.40\pm 0.02}$ (row 6 of Table~3). While the two power-law slopes are statistically consistent, $\sigma_c$ for FRs in \atlas drops off rapidly at $L_K \la 10^{10.75} L_{K,\odot}$, and
the scatter in $\sigma_c$ (at a given $L_K$) increases dramatically at low $L_K$.

\subsection{Choice of Velocity Dispersion Aperture}

The FJ relation has been studied using velocity dispersions measured within various aperture sizes.  Our results above used the central velocity dispersions measured within $\sim 1$ kpc for both MASSIVE and \atlas galaxies.
To check that our results are not sensitive to the exact aperture used to measure velocity dispersions, we replace $\sigma_c$ with $\sigma_e$, the velocity dispersion inferred within approximately one half-light radius of each galaxy. For MASSIVE galaxies, $\sigma_e$ values are already reported in \citet{Vealeetal2018}.  We list both $\sigma_c$ and $\sigma_e$ from that work in Table~1.  For \atlas galaxies,
we use the velocity dispersions measured within an aperture defined by the half-light elliptical isophote in \citet{Cappellarietal2013b}.
Because the reported \atlas velocity dispersions include rotational velocities within each galaxy, we consider only SRs in both samples for which rotations have a negligible effect on $\sigma_e$.

The resulting $\sigma_e$-$L_K$ relation is shown in Fig.~\ref{fig:FJ_e_slow}.
We find that MASSIVE SRs follow $\sigma_e \propto L_K^{0.18\pm0.05}$ in comparison to $\sigma_c \propto L_K^{0.11\pm0.07}$, and the \atlas SRs follow $\sigma_e \propto L_K^{0.31\pm0.02}$ in comparison to $\sigma_c \propto L_K^{0.36\pm0.02}$. The normalization and scatter of each relation is given in the caption of Fig.~\ref{fig:FJ_e_slow}. 
Overall, we find the FJ relation 
to be consistent (within $\sim 1\sigma$) between the two choices of velocity dispersion.

\subsection{Discussion}

The slope for the full MASSIVE sample ($\sigma_c \propto L_K^{0.19\pm0.06}$) is consistent with previous $K$-band studies of the FJ relation from both \citet{Pahre1998} and \citet{LaBarberaetal2010}, where the slopes are found to be 0.24 and 0.22, respectively. \citet{LaBarberaetal2010} find that the slope of the FJ relation is relatively constant over a range of wavelength bands. In the r-band, \citet{Bernardi2007} report a slope of 0.25 for SDSS early-type galaxies, consistent with our measured $K$-band slope in MASSIVE.

Curvature in the FJ relation has been suggested by \citet{OegerleHoessel1991}, with many more recent
studies supporting this trend. \citet{Lauer2007a} find $L\sim\sigma^{6.5\pm1.3}$ for core galaxies, compared to $L\sim\sigma^{2.6\pm0.3}$ for power-law galaxies which are typically less massive. Curvature was also reported in the FJ relation for early-type galaxies in the SDSS in \citet{HydeBernardi2009}. In this work, similar curvature is observed in the FJ relation. However, this curvature is particularly prominent within the SR sample. In this case, less luminous SRs in \atlas exhibit a $\sigma_c \propto L_K^{(0.36\pm0.03}$ relationship, compared to $\sigma_c \propto L_K^{(0.11\pm0.07}$ for the MASSIVE SRs. The relationship for MASSIVE SRs is consistent with a slope of 0 to within 1.6$\sigma$. This stands in contrast to FRs, where the FJ slopes are consistent between MASSIVE and \atlas, though the slope for MASSIVE is highly uncertain due to the small number of high luminosity FRs. 

\begin{figure*}
\includegraphics[width=\textwidth]{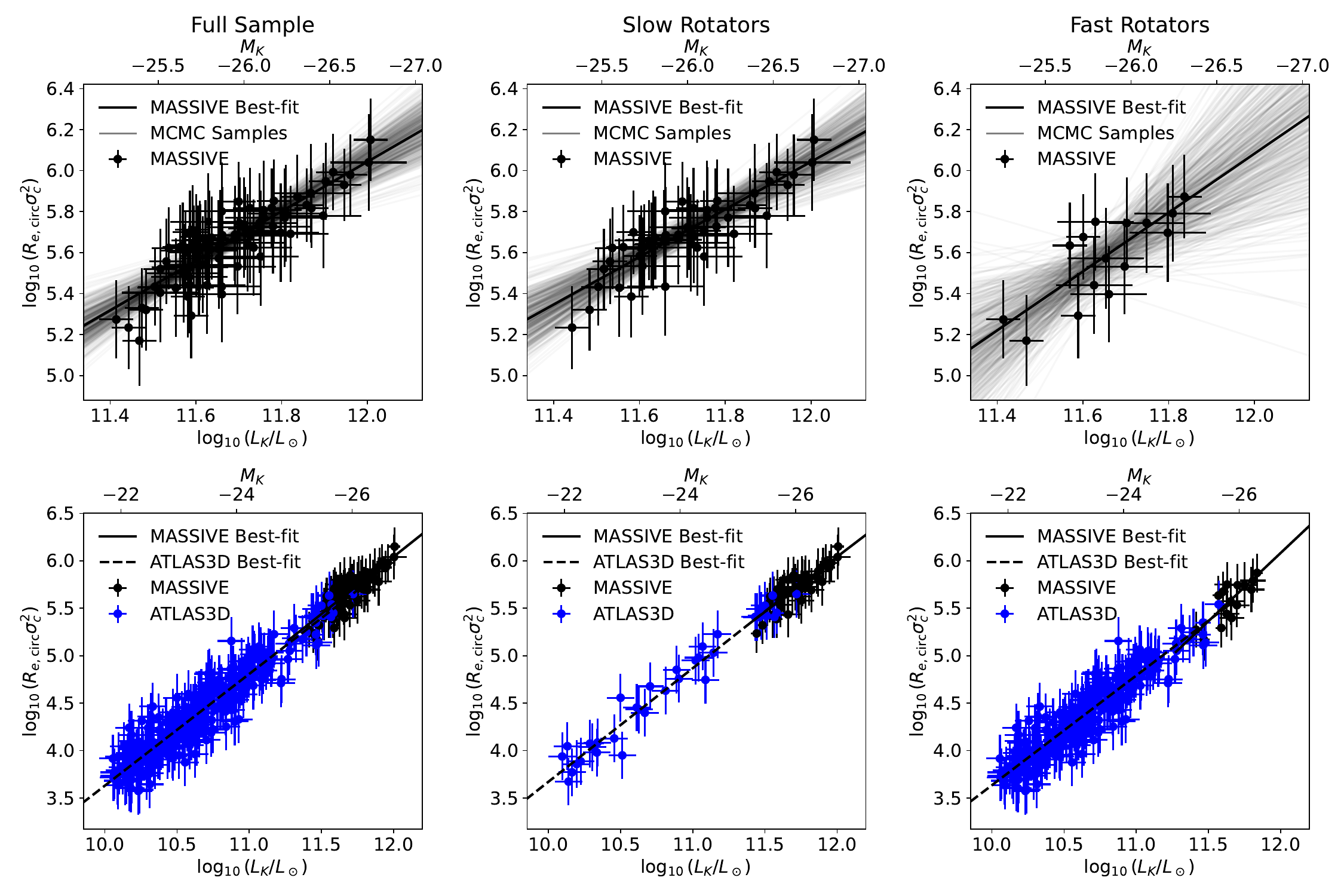}
\caption{ 
Relationship between dynamical mass $(M_\mathrm{dyn}\propto R_e \sigma_c^2)$ and total $K$-band luminosity for early-type galaxies in the MASSIVE survey (top) and in the MASSIVE and \atlas surveys together (bottom). The full samples (left column) are separated into slow rotators (middle) and fast rotators (right). The best-fit linear relations to various samples are summarized in
Table~\ref{tab:virial_c_params}. In the top row, the grey lines represent MCMC draws from the posterior distribution over the parameters describing the linear relations for MASSIVE galaxies.
The curvatures in the size-luminosity and FJ relations cancel, leading to a relation that is well described by a single power law: $R_e \sigma_c^2 \propto L^{b_{\rm vir}}$, with $b_{\rm vir} \approx 1.2$.}
\label{fig:virial_c}
\end{figure*}

FJ relation studies focused on BCGs find $L\sim\sigma^{5.32\pm0.37}$
\citep{vonderLindenetal2007} and $\sigma\sim L^{0.16\pm0.01}$ \citep{Samiretal2020}, suggesting that BCGs may have velocity dispersions that increase more slowly with luminosity than non-BCGs of comparable masses. Such a difference in slope, together with the increasing fraction of BCGs at large luminosities would lead to an overall curvature in the FJ relation. With $K$-band photometry from 2MASS, \citet{Batcheldoretal2007} analyze a sample of BCGs and do not find a significant difference in the $\sigma-L$ distribution of BCGs when compared to other ellipticals. However, \citet{Lauer2007a} suggest that this may be due to 2MASS being
insufficiently deep to recover accurate total magnitudes. In contrast, \citet{vonderLindenetal2007}
find a flatter $\sigma-L$ relation for BCGs than non-BCGs using 2MASS $K$-band photometry, consistent with our results given the larger fraction of BCGs in MASSIVE compared to \atlas. As with the SL relation, the MASSIVE sample does not contain enough satellite galaxies to determine whether BCG and satellite galaxies exhibit distinct FJ relations. Such changes in slope have also been predicted in simulations for BCGs. These galaxies may form through dissipationless mergers along orbits that are preferentially radial, leading to a flattening of the $\sigma-L$ relation~\citep[e.g.,][]{Boylan-Kolchin2006}.

\section{Dynamical Mass-Luminosity Relation}
\label{sec:virial}

The quantity $R_e \sigma^2$ is a proxy for the dynamical mass of a galaxy. The ratio between $R_e \sigma^2$ and total luminosity should therefore trace the dynamical mass-to-light ratio. This can be studied using a dynamical mass-luminosity power law relation
\begin{equation}
\label{eq:virial_c}
    \log_{10} \left({\frac{R_{e,\mathrm{circ}}\;\sigma_c^2}{\mathrm{1\;kpc\; km^2\; s^{-2}}}} \right)
    = b_\mathrm{vir,c} \log_{10} \left( {\frac{L_K}{10^{11.5}\;L\odot}} \right) +a_\mathrm{vir,c} 
\end{equation}
with a vertical scatter of $\epsilon_\mathrm{vir,c}$ about this relation. When using $\sigma_e$ in place of $\sigma_c$, we define:
\begin{equation}
\label{eq:virial_e}
    \log_{10} \left( {\frac{R_{e,\mathrm{circ}}\;\sigma_e^2}{\mathrm{1\;kpc\; km^2\; s^{-2}}}} \right) = b_\mathrm{vir,e} \log_{10} \left( {\frac{L_K}{10^{11.5}\;L\odot}} \right) + a_\mathrm{vir,e} 
\end{equation}
with a vertical scatter of $\epsilon_\mathrm{vir,e}$.

\begin{figure}
	\includegraphics[width=0.9\linewidth]{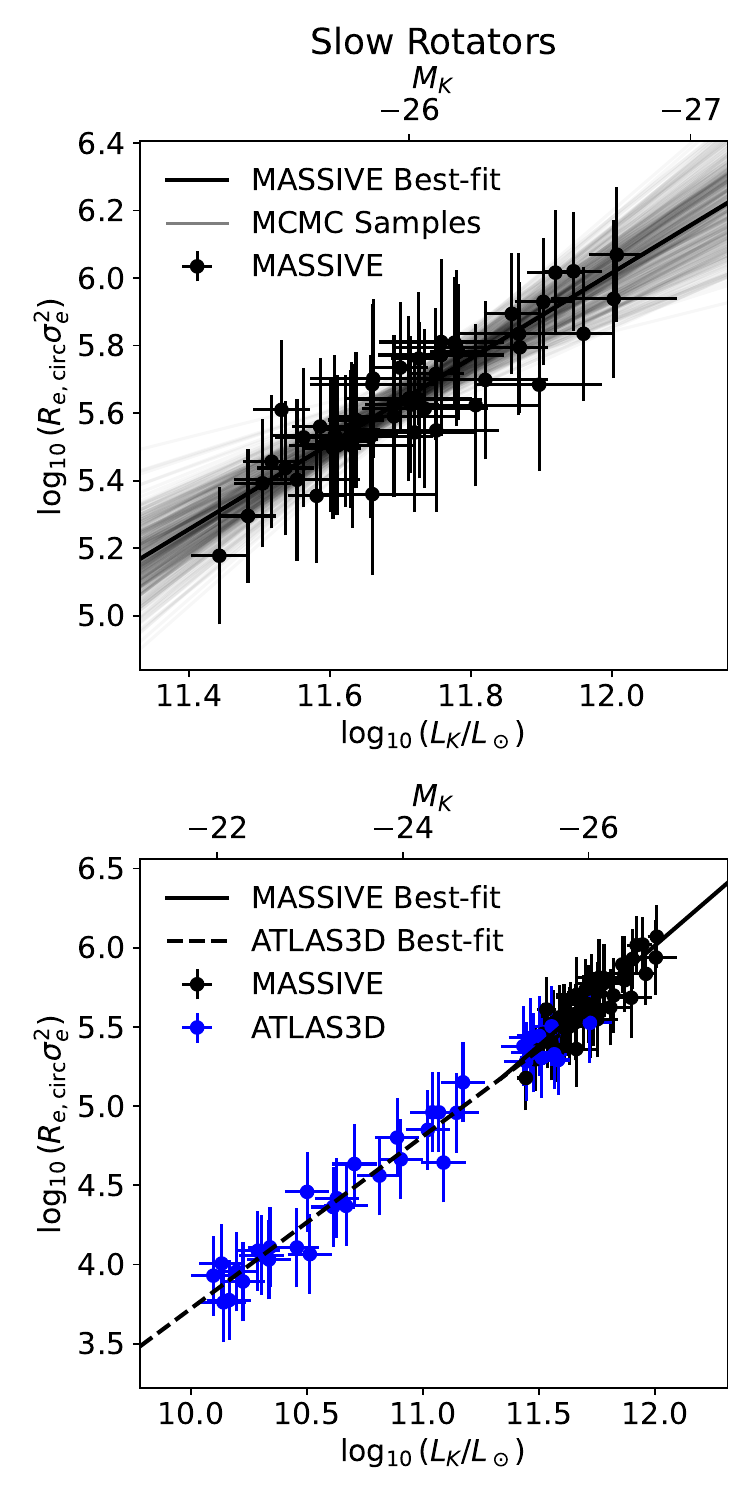}
    \caption{Same as the middle panels for slow rotators in Fig.~\ref{fig:virial_c}, except the central velocity dispersion $\sigma_c$ is replaced with $\sigma_e$ measured within the half-light radius. The best-fit line for MASSIVE is given
    by the parameters $a_\mathrm{vir,e}=5.41\pm0.06$, $b_\mathrm{vir,e}=1.2\pm0.2$, and
    $\epsilon_\mathrm{vir,e}=0.04^{+0.03}_{-0.02}$. There is little change to the relation between MASSIVE and \atlas. The best-fit line for \atlas is given by the
    parameters $a_\mathrm{vir,e}=5.36^{+0.07}_{-0.06}$, $b_\mathrm{vir,e}=1.09^{+0.09}_{-0.08}$, and
    $\epsilon_\mathrm{vir,e}=0.06^{+0.04}_{-0.03}$.}
    \label{fig:virial_e}
\end{figure}

We find that galaxies in the MASSIVE and \atlas surveys lie on a single
tight power-law relation in
$R_{e,\mathrm{circ}} \sigma_c^2$ and $L_K$, as shown in Fig.~\ref{fig:virial_c}.
The best-fit power-law relations for various galaxy samples are given in Table~\ref{tab:virial_c_params}.
Regardless of galaxy properties (e.g., luminosity or spin), the power-law relations are all consistent with 
$R_{e,\mathrm{circ}} \sigma_c^2 \propto L_K^{b_{\rm vir}}$, where $b_{\rm vir} \sim 1.2$.
In this sense, the curvatures in the SL and FJ relations cancel to give a single power law. 
This uniformity is evidence that the observed curvature is not merely a consequence of the different definitions of $\sigma_c$ and the corrections applied to $R_{e,\mathrm{circ}}$ and $L_K$ for \atlas. 

When $\sigma_c$ is replaced with $\sigma_e$, we find a very similar power-law relation for the SRs (see Fig.~\ref{fig:virial_e}). 
Only the SRs are presented due to the differing definitions of $\sigma_e$ between MASSIVE and \atlas. Again, the curvatures in the SL and FJ relations have cancelled out when $R_e \sigma_e^2$ is plotted against $L_K$. This suggests that the observed curvature in the FJ and SL relations is not merely a consequence of the different definitions of $\sigma_e$ between MASSIVE and \atlas. The corresponding fit values are summarized in the caption of Fig.~\ref{fig:virial_e}.

Deviations from the scalar virial theorem have been widely studied in the form of a ``tilt" in the fundamental plane for elliptical galaxies \citep[e.g.,][]{Pahre1998}. Assuming the galaxies have virialized, this tilt can be attributed to two possible causes: a systematic variation of the mass-to-light ratio, or a systematic variation of the structure or homology of elliptical galaxies.
In our sample, we find scalings that are consistent with $R_{e,\mathrm{circ}} \sigma^2\propto L^{1.2}$. Interpreting $R_{e,\mathrm{circ}} \sigma^2$ as a measure of dynamical mass suggests a scaling of the mass-to-light ratio of $(M/L)_\mathrm{vir}\propto L^{0.2}$. This is consistent with a scaling of $(M/L)_\mathrm{vir}\propto L^{0.2}$ for elliptical galaxies reported
in previous work \citep[e.g.,][]{Faber1987,Magorrian1998}.  \citet{Trujilloetal2004} attribute this same scaling entirely to non-homology among the SB profiles of ellipticals. \citet{Cappellarietal2006}, on the other hand, suggest that non-homology plays a negligible role in the tilt of the fundamental plane. Our results are consistent with prior measurements of the tilt of the fundamental plane, but more detailed modeling would be needed to distinguish between these two scenarios.

\begin{table}
    \begin{tabular}{|l|lll|}
    \hline
    Sample & $a_\mathrm{vir,c}$ & $b_\mathrm{vir,c}$ & $\epsilon_\mathrm{vir,c}$ \\ \hline
    MASSIVE (full) & $5.45\pm0.04$ & $1.3\pm0.2$ & $0.03\pm0.02$ \\
    \atlas\ (full) & $5.39\pm0.04$ & $1.17^{+0.05}_{-0.04}$ & $0.026^{+0.017}_{-0.013}$ \\ \hline
    MASSIVE (SRs) & $5.48\pm0.05$ & $1.2\pm0.2$ & $0.04^{+0.03}_{-0.02}$ \\
    \atlas (SRs) & $5.47^{+0.08}_{-0.07}$ & $1.19\pm0.09$ & $0.06^{+0.04}_{-0.03}$ \\ \hline
    MASSIVE (FRs) & $5.38^{+0.10}_{-0.11}$ & $1.4\pm0.6$ & $0.10^{+0.07}_{-0.05}$ \\
    \atlas (FRs) & $5.37^{+0.04}_{-0.05}$ & $1.15\pm0.05$ & $0.023^{+0.016}_{-0.012}$ \\ \hline
    \end{tabular}
    \caption{Fit parameters for the dynamical mass-luminosity relation, as defined by equation~\ref{eq:virial_c}.
    The galaxy subsamples are the same as in Table~\ref{tab:SL_params}.
        }
    \label{tab:virial_c_params}
\end{table}

\section{Conclusions}

We have presented CFHT WIRCam data for 98 galaxies in the MASSIVE survey, and measured the $K$-band total luminosities and half-light radii. The updated luminosities are systematically
brighter by about 0.29~mag than those from the 2MASS XSC.
The half-light radii are systematically larger than those from 2MASS by about 18\%.

Using these measured values, we study the SL and FJ relations for the MASSIVE galaxies. For the SL
relation, we find $R_e\sim L_K^{0.93\pm0.10}$. This is consistent with prior studies, and indicates a
significant steepening of the SL relation at high luminosities. For SRs alone, \atlas and MASSIVE
combine to form a tight continuous sequence in the $R_e-L$ plane, but once again, there is a significant
steepening of the SL relation at luminosities $L_K\gtrsim10^{11.5}~L_{K,\odot}$.

For the FJ relation, we find $L\sim\sigma_c^{5}$ for the full MASSIVE sample. However, when we
consider the SRs alone, the power law steepens to $L\sim\sigma_c^{9}$. Consistent with other
studies, we find $\sigma$ flattens as a function of luminosity for the most massive galaxies.
This is likely related to the prevalence of central cores in these massive galaxies.

The curvature in these relations is further evidence for a picture in which the most
luminous elliptical galaxies grow mainly through dissipationless ``dry'' mergers, with dissipation
playing a larger role for less luminous galaxies. Even within a sample of only slow
rotators, there is a significant change in the slopes of the SL and FJ relations, suggesting
different formation histories for low and high luminosity slow rotators. Prior studies have suggested distinct populations of low and high mass SR elliptical galaxies based
on other evidence. \citet{Krajnovicetal2013} showed that while the most massive ellipticals are
cored, some less massive SRs can be core-less. Thus, the difference between low-mass core-less and
high-mass cored galaxies is not simply related to angular momentum. The increased prevalence of
cores among SRs occurs around a characteristic stellar dynamical mass of $2\times10^{11}~M_\odot$,
which corresponds to an absolute $K$ magnitude $\sim-24.9$ mag using the stellar mass
relation from \citet{Cappellari2013} together with the correction between 2MASS and CFHT magnitudes
reported in Section~\ref{sec:photometric_params}. This characteristic luminosity is qualitatively
consistent with what we observe in both the SL and FJ relations. 

Scaling relations between SMBH masses, \mbh, and the velocity dispersions or luminosities of the host galaxies are often used to predict \mbh\ in galaxies where it cannot be
measured directly. 
If $\sigma$ and $L$ are related by a simple power-law, then power-law 
fits to the $M_\mathrm{BH}-\sigma$ and
$M_\mathrm{BH}-L$ relations
should be consistent with one another. However, if there is curvature in
the relationship between $\sigma$ and $L$, power-law fits to the two scaling relations can result in different predictions for \mbh\ \citep{Lauer2007a}. There is evidence that the $M_\mathrm{BH}-\sigma$
relation underpredicts the highest SMBH masses~\citep[e.g.,][]{McConnellMa2013,Thomasetal2016},
although this may be due to differing $M_\mathrm{BH}-\sigma$ relations for cored and non-cored galaxies
\citep{Sahuetal2019}, with a predominance of cored galaxies at the highest luminosities. 
The MASSIVE survey will provide further explorations of SMBH scaling relations as we increase the number of dynamical black hole detections and mass measurements at the highest masses \citep{Liepoldetal2020, Quennevilleetal2022, Pilawaetal2022}.

\section*{Acknowledgements}
We thank James Schombert for his assistance with the ARCHANGEL pipeline. This work is based on observations obtained with WIRCam, a joint project of CFHT, Taiwan, Korea, Canada, France, at the Canada-France-Hawaii Telescope (CFHT) which is operated by the National Research Council (NRC) of Canada, the Institut National des Sciences de l'Univers of the Centre National de la Recherche Scientifique of France, and the University of Hawaii.
We thank the staff of CFHT for their outstanding and always good-natured support during the course of this project.
This work is supported in part by NSF AST-1815417, AST-1817100 and AST-2206219.
J.P.B.\ is supported by NOIRLab, which is managed by the Association of Universities for Research in Astronomy (AURA) under a cooperative agreement with the National Science Foundation.
C.-P.M. acknowledges support from the Heising-Simons Foundation and the Miller Institute for Basic Research in Science. 

\section*{Data Availability}
Data available on reasonable request.

%%%%%%%%%%%%%%%%%%%%%%%%%%%%%%%%%%%%%%%%%%%%%%%%%%

%%%%%%%%%%%%%%%%%%%% REFERENCES %%%%%%%%%%%%%%%%%%

% The best way to enter references is to use BibTeX:

\bibliographystyle{mnras}
\bibliography{massive_cfht}
%\bibliography{bibliography} % if your bibtex file is called example.bib

%%%%%%%%%%%%%%%%%%%%%%%%%%%%%%%%%%%%%%%%%%%%%%%%%%

%%%%%%%%%%%%%%%%% APPENDICES %%%%%%%%%%%%%%%%%%%%%

%\appendix

%%%%%%%%%%%%%%%%%%%%%%%%%%%%%%%%%%%%%%%%%%%%%%%%%%

% Don't change these lines
\bsp	% typesetting comment
\label{lastpage}
\end{document}

%% file: tables/main_table.tex
\begin{table*}
	\caption{(1) Galaxy name. (2) Angular semi-major axis of the half-light elliptical isophote
          from $K$-band CFHT imaging. (3) Angular semi-major axis of the half-light elliptical isophote
          from 2MASS in $K$-band (\textit{k\_r\_eff}). (4) Ellipticity of the half-light elliptical
          isophote from $K$-band CFHT imaging. (5)~Ellipticity of the 3-$\sigma$ elliptical isophote
          from 2MASS in the combined $J$, $H$, $K$ band image (1-\textit{sup\_ba}). (6)~Total $K$-band apparent
          magnitude from CFHT imaging. (7)~Total $K$-band apparent magnitude from 2MASS. (8) Distance,
          as reported in \citet{Jensenetal2021} where available (indicated by *), otherwise from
          \citet{Maetal2014}. (9) Total $K$-band absolute magnitude based on CFHT
          imaging. (10)~Semi-major axis of the half-light elliptical isophote from $K$-band CFHT
          imaging, converted to physical units using the adopted distance. (11)~Geometric radius of
          the half-light elliptical isophote from $K$-band CFHT imaging, in physical
          units. (12)~Velocity dispersion within the half-light radius as reported by
          \citet{Vealeetal2018}, where the radius comes from the NSA where available or 2MASS
          corrected to agree with NSA on average. (13)~Central velocity dispersion
          \citep{Vealeetal2018}. 
          (14) Slow (``S") or fast (``F") rotator \citep{Vealeetal2018}.
          \protect\\ \protect\begin{small}$\dagger$\quad NGC0545 and NGC0547 are a close galaxy pair with overlapping isophotes and reliable photometric parameters could not be measured.\protect\end{small}
	}
	\label{tab:param_vals}
	\begin{tabular}{lccccccccccccc}
		\hline
		Name & $R_e^\mathrm{}$ & $R_e^\mathrm{2MASS}$ & $\varepsilon^\mathrm{}$ & $\varepsilon^\mathrm{2MASS}$ & $K^\mathrm{}$ & $K^\mathrm{2MASS}$ & $D$ & $M_K^\mathrm{}$ & $R_e^\mathrm{}$ & $R_{e,\mathrm{circ}}^\mathrm{}$ & $\sigma_{e}^\mathrm{}$ & $\sigma_{c}^\mathrm{}$ & Rot\\
		~ & ($''$) & ($''$) & ~ & ~ & (mag) & (mag) & (Mpc) & (mag) & (kpc) & (kpc) & (km/s) & (km/s) & \\
		(1) & (2) & (3) & (4) & (5) & (6) & (7) & (8) & (9) & (10) & (11) & (12) & (13) & (14) \\
		\hline
		NGC0057 & 16.91 & 14.65 & 0.14 & 0.2 & 8.4 & 8.68 & 66.9$^{*}$ & -25.75 & 5.48 & 5.07 & 251 & 289 & S \\
  
        NGC0080 & 21.04 & 16.39 & 0.11 & 0.08 & 8.64 & 8.92 & 81.9 & -25.95 & 8.35 & 7.9 & 222 & 248 & S \\
 
		NGC0128 & 14.99 & 19.22 & 0.5 & 0.7 & 8.35 & 8.52 & 59.3 & -25.53 & 4.31 & 3.04 & ~ & ~ & \\
 
		NGC0227 & 13.65 & 10.0 & 0.35 & 0.24 & 8.73 & 9.09 & 75.9 & -25.68 & 5.02 & 4.05 & ~ & ~ & \\
 
		NGC0315 & 25.7 & 22.15 & 0.26 & 0.24 & 7.71 & 7.96 & 68.1$^{*}$ & -26.48 & 8.48 & 7.31 & 341 & 348 & S\\
 
		NGC0383 & 22.11 & 16.23 & 0.13 & 0.16 & 8.15 & 8.48 & 66.1$^{*}$ & -25.97 & 7.09 & 6.59 & 257 & 290 & F\\
 
		NGC0393 & 14.5 & 11.93 & 0.17 & 0.18 & 8.93 & 9.23 & 85.7 & -25.75 & 6.03 & 5.49 & ~ & ~ & \\
 
		NGC0410 & 21.11 & 19.18 & 0.26 & 0.26 & 8.18 & 8.38 & 61.3$^{*}$ & -25.78 & 6.27 & 5.39 & 247 & 291 & S \\
 
		NGC0467 & 19.49 & 14.84 & 0.05 & 0.08 & 8.65 & 9.01 & 75.8 & -25.76 & 7.16 & 6.98 & ~ & ~ & \\
 
		PGC04829 & 9.01 & 8.8 & 0.26 & 0.34 & 9.46 & 9.73 & 99.0 & -25.58 & 4.33 & 3.71 & ~ & ~ & \\
 
		NGC0499 & 13.2 & 14.52 & 0.36 & 0.36 & 8.64 & 8.74 & 69.8 & -25.60 & 4.47 & 3.57 & 266 & 274 & S\\
 
		NGC0507 & 31.75 & 23.98 & 0.15 & 0.08 & 7.94 & 8.3 & 61.7$^{*}$ & -26.03 & 9.5 & 8.74 & 257 & 274 & S\\
 
		NGC0533 & 29.63 & 25.16 & 0.25 & 0.24 & 8.15 & 8.44 & 72.8$^{*}$ & -26.17 & 10.46 & 9.08 & 258 & 280 & S\\
 
		NGC0545$^\dagger$ & ~ & ~ & ~ & ~ & ~ & ~ & 79.0 & ~ & ~ & ~ & 231 & 249 & S\\
  
		NGC0547$^\dagger$ & ~ & 29.58 & ~ & 0.28 & ~ & 8.49 & 75.5 & ~ & ~ & ~ & 232 & 259 & S\\
  
		NGC0665 & 13.27 & 13.37 & 0.24 & 0.38 & 8.61 & 8.88 & 62.3$^{*}$ & -25.39 & 4.01 & 3.49 & 164 & 206 & F\\
 
		UGC01332 & 18.79 & 15.39 & 0.24 & 0.3 & 9.17 & 9.48 & 99.2 & -25.87 & 9.04 & 7.87 & 253 & 248 & S\\
 
		NGC0708 & 34.04 & 30.59 & 0.3 & 0.4 & 8.42 & 8.57 & 61.5$^{*}$ & -25.55 & 10.15 & 8.49 & 219 & 206 & S\\
 
		UGC01389 & 13.03 & 9.84 & 0.14 & 0.34 & 9.26 & 9.63 & 99.2 & -25.78 & 6.27 & 5.83 & ~ & ~ &\\
 
		NGC0741 & 25.17 & 22.8 & 0.18 & 0.36 & 8.07 & 8.3 & 69.6$^{*}$ & -26.16 & 8.49 & 7.71 & 289 & 292 & S\\
 
		NGC0777 & 16.69 & 15.36 & 0.18 & 0.1 & 8.17 & 8.37 & 68.0$^{*}$ & -26.01 & 5.5 & 4.98 & 291 & 324 & S\\
 
		NGC0890 & 23.14 & 20.5 & 0.39 & 0.38 & 7.99 & 8.24 & 45.6$^{*}$ & -25.33 & 5.11 & 4.0 & 194 & 207 & S \\
 
		NGC0910 & 19.83 & 14.05 & 0.08 & 0.16 & 8.82 & 9.2 & 77.9$^{*}$ & -25.66 & 7.49 & 7.17 & 219 & 236 & S \\
 
		NGC0997 & 11.66 & 10.13 & 0.14 & 0.18 & 9.14 & 9.42 & 90.4 & -25.68 & 5.11 & 4.73 & 215 & 267 & F  \\
 
		NGC1016 & 19.7 & 18.48 & 0.05 & 0.08 & 8.29 & 8.58 & 85.9$^{*}$ & -26.39 & 8.2 & 8.01 & 279 & 286 & S \\
 
		NGC1060 & 19.9 & 17.7 & 0.23 & 0.14 & 7.92 & 8.2 & 53.8$^{*}$ & -25.79 & 5.19 & 4.54 & 271 & 310 & S\\
 
		NGC1066 & 19.79 & 18.56 & 0.2 & 0.2 & 8.72 & 8.89 & 67.4 & -25.48 & 6.47 & 5.78 & ~ & ~ & \\
 
		NGC1132 & 24.42 & 18.12 & 0.36 & 0.34 & 8.97 & 9.26 & 97.6 & -26.0 & 11.56 & 9.25 & 218 & 239 & S \\
 
		NGC1129 & 41.12 & 29.49 & 0.22 & 0.2 & 7.77 & 8.24 & 66.2$^{*}$ & -26.37 & 13.2 & 11.69 & 259 & 241 & S\\
 
		NGC1167 & 22.35 & 20.88 & 0.17 & 0.22 & 8.45 & 8.64 & 53.7$^{*}$ & -25.26 & 5.82 & 5.31 & 172 & 188 & F \\
 
		NGC1226 & 15.95 & 13.83 & 0.16 & 0.18 & 8.92 & 9.21 & 85.7 & -25.8 & 6.63 & 6.08 & 229 & 274 & S\\
 
		IC0310 & 13.41 & 12.08 & 0.05 & 0.04 & 8.9 & 9.14 & 71.0$^{*}$ & -25.41 & 4.62 & 4.5 & 205 & 218 & S \\
 
		NGC1272 & 25.77 & 20.89 & 0.04 & 0.02 & 8.34 & 8.69 & 71.0$^{*}$ & -25.97 & 8.87 & 8.69 & 250 & 285 & S \\
 
		UGC02783 & 9.44 & 8.97 & 0.11 & 0.16 & 9.07 & 9.27 & 85.8 & -25.65 & 3.93 & 3.7 & 266 & 292 & S \\
 
		NGC1453 & 19.76 & 17.26 & 0.19 & 0.14 & 7.93 & 8.12 & 51.2$^{*}$ & -25.64 & 4.91 & 4.43 & 272 & 312 & F\\
 
		NGC1497 & 13.39 & 13.27 & 0.34 & 0.4 & 9.27 & 9.48 & 87.8 & -25.51 & 5.7 & 4.63 & 190 & 234 & F \\
 
		NGC1600 & 27.93 & 24.15 & 0.33 & 0.26 & 7.67 & 8.04 & 71.7$^{*}$ & -26.62 & 9.71 & 7.97 & 293 & 346 & S\\
 
		NGC1573 & 17.86 & 17.14 & 0.29 & 0.34 & 8.33 & 8.56 & 63.5$^{*}$ & -25.72 & 5.5 & 4.62 & 264 & 288 & S \\
 
		NGC1684 & 23.02 & 17.01 & 0.29 & 0.24 & 8.31 & 8.69 & 62.8$^{*}$ & -25.7 & 7.01 & 5.91 & 262 & 295 & S \\
 
		NGC1700 & 16.34 & 14.89 & 0.27 & 0.28 & 7.91 & 8.09 & 52.2$^{*}$ & -25.69 & 4.13 & 3.52 & 223 & 236 & F \\
 
		NGC2208 & 21.86 & 17.16 & 0.43 & 0.32 & 8.63 & 9.04 & 84.1 & -26.04 & 8.91 & 6.7 & 255 & 268 & S \\
 
		NGC2256 & 26.76 & 22.62 & 0.19 & 0.2 & 8.37 & 8.67 & 79.4 & -26.17 & 10.3 & 9.27 & 259 & 240 & S \\
 
		NGC2274 & 18.0 & 15.8 & 0.12 & 0.1 & 8.42 & 8.68 & 69.1$^{*}$ & -25.81 & 6.03 & 5.66 & 259 & 288 & S \\
 
		NGC2258 & 19.82 & 19.92 & 0.21 & 0.24 & 8.09 & 8.23 & 57.0$^{*}$ & -25.73 & 5.48 & 4.86 & 254 & 293 & S \\
 
		NGC2320 & 16.27 & 12.69 & 0.42 & 0.3 & 8.53 & 8.85 & 89.4 & -26.24 & 7.05 & 5.35 & 298 & 340 & F \\
 
		UGC03683 & 14.11 & 13.07 & 0.21 & 0.26 & 8.94 & 9.15 & 85.1 & -25.74 & 5.82 & 5.18 & 257 & 257 & S \\
 
		NGC2332 & 13.51 & 11.01 & 0.36 & 0.34 & 9.1 & 9.39 & 89.4 & -25.68 & 5.86 & 4.69 & 224 & 254 & S \\
 
		NGC2340 & 32.58 & 26.65 & 0.44 & 0.44 & 8.44 & 8.88 & 79.9$^{*}$ & -26.1 & 12.62 & 9.42 & 235 & 232 & S \\
 
		UGC03894 & 14.3 & 13.05 & 0.1 & 0.12 & 9.17 & 9.37 & 97.2 & -25.79 & 6.74 & 6.38 & 255 & 297 & F \\
 
		NGC2418 & 14.46 & 11.87 & 0.21 & 0.12 & 8.68 & 8.94 & 74.1 & -25.68 & 5.19 & 4.61 & 217 & 245 & F \\
 
		NGC2456 & 15.27 & 12.15 & 0.25 & 0.24 & 9.5 & 9.83 & 107.3 & -25.67 & 7.94 & 6.86 & ~ & ~ & \\
 
		NGC2492 & 10.82 & 9.43 & 0.22 & 0.16 & 9.35 & 9.6 & 97.8 & -25.62 & 5.13 & 4.53 & ~ & ~ & \\
 
		NGC2513 & 17.97 & 15.58 & 0.19 & 0.2 & 8.47 & 8.74 & 71.1$^{*}$ & -25.8 & 6.2 & 5.57 & 253 & 280 & S

	\end{tabular}
\end{table*}

\begin{table*}
	\begin{tabular}{lccccccccccccc}
		\hline
		Name & $R_e^\mathrm{}$ & $R_e^\mathrm{2MASS}$ & $\varepsilon^\mathrm{}$ & $\varepsilon^\mathrm{2MASS}$ & $K^\mathrm{}$ & $K^\mathrm{2MASS}$ & $D$ & $M_K^\mathrm{}$ & $R_e^\mathrm{}$ & $R_{e,\mathrm{circ}}^\mathrm{}$ & $\sigma_{e}^\mathrm{}$ & $\sigma_{c}^\mathrm{}$ & Rot \\
		~ & (") & (") & ~ & ~ & (mag) & (mag) & (Mpc) & (mag) & (kpc) & (kpc) & (km/s) & (km/s) & \\
		(1) & (2) & (3) & (4) & (5) & (6) & (7) & (8) & (9) & (10) & (11) & (12) & (13) & (14)\\
		\hline

        NGC2672 & 20.86 & 18.8 & 0.15 & 0.28 & 8.16 & 8.35 & 66.2$^{*}$ & -25.95 & 6.7 & 6.17 & 262 & 273 & S \\
   
		NGC2693 & 15.04 & 15.92 & 0.27 & 0.26 & 8.54 & 8.6 & 71.0$^{*}$ & -25.72 & 5.18 & 4.43 & 296 & 327 & F \\
		
		NGC2783 & 24.47 & 15.09 & 0.4 & 0.44 & 8.92 & 9.32 & 101.4 & -26.11 & 12.03 & 9.28 & 264 & 252 & S \\
 
		NGC2832 & 23.12 & 20.08 & 0.25 & 0.18 & 8.39 & 8.7 & 105.2 & -26.73 & 11.79 & 10.25 & 291 & 327 & S \\
 
		NGC2892 & 15.57 & 13.09 & 0.03 & 0.08 & 8.99 & 9.35 & 101.1 & -26.06 & 7.63 & 7.5 & 234 & 237 & S \\
 
		NGC2918 & 11.0 & 9.92 & 0.32 & 0.22 & 9.38 & 9.57 & 102.3 & -25.67 & 5.46 & 4.51 & ~ & ~ & \\
 
		NGC3158 & 20.44 & 15.45 & 0.18 & 0.16 & 8.5 & 8.8 & 91.5$^{*}$ & -26.31 & 9.07 & 8.23 & 289 & 301 & F \\
 
		NGC3209 & 14.04 & 10.3 & 0.26 & 0.24 & 9.02 & 9.34 & 94.6 & -25.86 & 6.44 & 5.55 & 247 & 288 & S \\
 
		NGC3332 & 16.99 & 13.2 & 0.14 & 0.08 & 9.08 & 9.37 & 89.1 & -25.68 & 7.34 & 6.82 & ~ & ~ & \\
 
		NGC3343 & ~ & 12.68 & ~ & 0.32 & ~ & 9.57 & 93.8 & ~ & ~ & ~ & ~ & ~ & \\
  
		NGC3462 & 12.03 & 11.27 & 0.25 & 0.2 & 9.12 & 9.37 & 99.2 & -25.87 & 5.79 & 5.0 & 214 & 233 & S \\
 
		NGC3562 & 13.9 & 9.08 & 0.2 & 0.16 & 8.94 & 9.38 & 101.0 & -26.1 & 6.81 & 6.09 & 241 & 250 & S \\
 
		NGC3615 & 13.05 & 10.08 & 0.34 & 0.34 & 9.18 & 9.45 & 101.2 & -25.85 & 6.4 & 5.2 & 232 & 268 & F \\
 
		NGC3805 & ~ & 9.4 & ~ & 0.26 & ~ & 9.3 & 99.4 & ~ & ~ & ~ & 225 & 266 & F \\
  
		NGC3816 & ~ & 13.76 & ~ & 0.4 & ~ & 9.6 & 99.4 & ~ & ~ & ~ & 191 & 212 & S \\ 
  
		NGC3842 & ~ & 15.62 & ~ & 0.18 & ~ & 9.08 & 87.5 & ~ & ~ & ~ & 231 & 262 & S \\
  
		NGC3862 & ~ & 11.12 & ~ & 0.08 & ~ & 9.49 & 99.4 & ~ & ~ & ~ & 228 & 248 & S \\
  
		NGC3937 & ~ & 12.53 & ~ & 0.24 & ~ & 9.42 & 101.2 & ~ & ~ & ~ & 243 & 292 & S \\
  
		NGC4055 & ~ & 8.63 & ~ & 0.12 & ~ & 9.76 & 107.2 & ~ & ~ & ~ & ~ & ~ & \\
  
		NGC4065 & ~ & 9.42 & ~ & 0.14 & ~ & 9.69 & 107.2 & ~ & ~ & ~ & ~ & ~ & \\
  
		NGC4066 & ~ & 10.36 & ~ & 0.16 & ~ & 9.81 & 107.2 & ~ & ~ & ~ & ~ & ~ & \\
  
		NGC4059 & ~ & 10.26 & ~ & 0.14 & ~ & 9.75 & 107.2 & ~ & ~ & ~ & ~ & ~ & \\
  
		NGC4073 & ~ & 25.27 & ~ & 0.34 & ~ & 8.49 & 85.0 & ~ & ~ & ~ & 292 & 316 & S \\
  
		NGC4213 & 15.59 & 13.26 & 0.16 & 0.24 & 9.36 & 9.61 & 101.6 & -25.69 & 7.68 & 7.05 & ~ & ~ & \\
 
		NGC4472 & ~ & 56.11 & ~ & 0.09 & ~ & 5.4 & 16.7 & ~ & ~ & ~ & 258 & 292 & F \\
  
		NGC4486 & ~ & 41.46 & ~ & 0.01 & ~ & 5.81 & 16.7 & ~ & ~ & ~ & ~ & ~ & S \\
  
		NGC4555 & 12.72 & 11.19 & 0.26 & 0.16 & 8.84 & 9.17 & 103.6 & -26.24 & 6.39 & 5.48 & 277 & 328 & S \\
   
		NGC4649 & ~ & 42.14 & ~ & 0.11 & ~ & 5.74 & 16.5 & ~ & ~ & ~ & ~ & ~ & \\
  
		NGC4816 & 21.49 & 15.63 & 0.2 & 0.14 & 9.28 & 9.71 & 99.1$^{*}$ & -25.71 & 10.33 & 9.22 & 207 & 217 & S \\
  
 		NGC4839 & ~ & 21.89 & ~ & 0.34 & ~ & 9.2 & 91.2 & ~ & ~ & ~ & 275 & 261 & S \\
   
		NGC4874 & 33.59 & 24.58 & 0.07 & 0.1 & 8.46 & 8.86 & 99.1$^{*}$ & -26.52 & 16.14 & 15.59 & 258 & 251 & S \\
 
		NGC4889 & 26.89 & 23.83 & 0.36 & 0.36 & 8.25 & 8.41 & 99.1$^{*}$ & -26.74 & 12.92 & 10.33 & 337 & 370 & S \\
 
		NGC4914 & 17.77 & 16.04 & 0.4 & 0.38 & 8.47 & 8.65 & 67.1$^{*}$ & -25.67 & 5.78 & 4.47 & 225 & 233 & S \\
 
		NGC5129 & 17.41 & 14.19 & 0.34 & 0.32 & 8.95 & 9.25 & 107.5 & -26.21 & 9.07 & 7.36 & 222 & 260 & F \\
 
		NGC5208 & 10.82 & 9.37 & 0.61 & 0.58 & 9.25 & 9.51 & 105.0 & -25.87 & 5.51 & 3.42 & 235 & 270 & F \\
 
		PGC47776 & ~ & 8.49 & ~ & 0.2 & ~ & 9.73 & 103.8 & ~ & ~ & ~ & ~ & ~ & \\
  
		NGC5252 & 15.28 & 13.16 & 0.58 & 0.5 & 9.48 & 9.77 & 103.8 & -25.61 & 7.69 & 4.99 & ~ & ~ & \\
 
		NGC5322 & 27.37 & 30.52 & 0.32 & 0.34 & 7.07 & 7.16 & 31.5$^{*}$ & -25.43 & 4.18 & 3.46 & 239 & 246 & S \\
 
		NGC5353 & ~ & 17.95 & ~ & 0.56 & ~ & 7.62 & 34.8 & ~ & ~ & ~ & 225 & 277 & F \\
  
		NGC5490 & 11.02 & 11.13 & 0.18 & 0.18 & 8.71 & 8.92 & 71.4$^{*}$ & -25.56 & 3.81 & 3.44 & 282 & 349 & S \\
  
 		NGC5557 & 19.8 & 17.38 & 0.16 & 0.14 & 7.78 & 8.08 & 49.2$^{*}$ & -25.68 & 4.72 & 4.34 & 223 & 279 & S \\
 
		IC1143 & 10.92 & 9.9 & 0.1 & 0.14 & 9.3 & 9.51 & 97.3 & -25.66 & 5.15 & 4.9 & ~ & ~ & \\
 
		UGC10097 & 10.18 & 8.5 & 0.3 & 0.26 & 9.07 & 9.38 & 91.5 & -25.74 & 4.52 & 3.77 & ~ & ~ & \\
 
		NGC6223 & 21.35 & 11.41 & 0.32 & 0.2 & 8.61 & 9.11 & 86.7 & -26.09 & 8.98 & 7.4 & 238 & 274 & F \\
 
		NGC6364 & 10.27 & 8.19 & 0.16 & 0.22 & 9.4 & 9.74 & 105.3 & -25.72 & 5.24 & 4.81 & ~ & ~ & \\
 
		NGC6375 & 12.5 & 12.17 & 0.13 & 0.1 & 9.16 & 9.41 & 95.8 & -25.78 & 5.81 & 5.4 & 187 & 226 & F \\
 
		UGC10918 & 18.24 & 12.18 & 0.17 & 0.14 & 8.79 & 9.31 & 100.2 & -26.27 & 8.86 & 8.05 & 249 & 247 & S \\
 
		NGC6442 & 12.62 & 9.75 & 0.21 & 0.12 & 9.13 & 9.59 & 98.0 & -25.86 & 6.0 & 5.35 & ~ & ~ & \\
 
		NGC6482 & 13.82 & 12.09 & 0.3 & 0.36 & 8.12 & 8.37 & 51.8$^{*}$ & -25.48 & 3.47 & 2.91 & 291 & 305 & S \\
 
		NGC6575 & 14.43 & 10.65 & 0.26 & 0.28 & 9.13 & 9.56 & 106.0 & -26.02 & 7.41 & 6.38 & 234 & 264 & S \\
 
		NGC7052 & 22.47 & 20.25 & 0.5 & 0.5 & 8.37 & 8.57 & 61.9$^{*}$ & -25.63 & 6.74 & 4.77 & 266 & 298 & S \\
 
		NGC7242 & 40.9 & 32.84 & 0.31 & 0.28 & 7.97 & 8.33 & 79.6$^{*}$ & -26.58 & 15.79 & 13.09 & 283 & 255 & S \\
 
		NGC7265 & 32.02 & 18.13 & 0.21 & 0.22 & 8.16 & 8.69 & 82.8 & -26.46 & 12.85 & 11.39 & 206 & 230 & S \\
 
		NGC7274 & 15.71 & 11.81 & 0.08 & 0.06 & 8.88 & 9.24 & 82.8 & -25.74 & 6.31 & 6.05 & 244 & 259 & S \\
 
		NGC7386 & 16.09 & 13.11 & 0.29 & 0.3 & 9.13 & 9.42 & 99.1 & -25.87 & 7.73 & 6.5 & 273 & 312 & S \\
 
		NGC7426 & 13.64 & 13.47 & 0.36 & 0.34 & 8.59 & 8.82 & 80.0 & -25.96 & 5.29 & 4.22 & 219 & 284 & F \\
 
		NGC7436 & 19.82 & 20.47 & 0.07 & 0.1 & 8.78 & 9.01 & 106.6 & -26.39 & 10.24 & 9.89 & 263 & 280 & S \\

 		NGC7550 & 16.9 & 12.9 & 0.08 & 0.1 & 8.5 & 8.91 & 72.7 & -25.85 & 5.96 & 5.71 & 224 & 270 & S \\
 
		NGC7556 & 23.03 & 18.93 & 0.24 & 0.2 & 8.96 & 9.25 & 103.0 & -26.11 & 11.5 & 10.02 & 243 & 253 & S \\
 
		NGC7618 & 13.4 & 11.26 & 0.32 & 0.28 & 8.72 & 9.04 & 76.3 & -25.76 & 4.96 & 4.07 & 265 & 292 & F \\
 
		NGC7619 & 23.7 & 16.39 & 0.22 & 0.18 & 7.68 & 8.03 & 46.6$^{*}$ & -25.68 & 5.35 & 4.73 & 277 & 325 & S \\
 
		NGC7626 & 21.92 & 20.66 & 0.15 & 0.12 & 7.85 & 8.03 & 46.6$^{*}$ & -25.51 & 4.95 & 4.57 & 250 & 269 & S \\
 
		NGC7681 & ~ & 12.39 & ~ & 0.22 & ~ & 9.22 & 96.8 & ~ & ~ & ~ & ~ & ~ & 
%	\end{tabular}
%\end{table*}
 	\end{tabular}
\end{table*}